# Nature Inspired Design Methodology for a Wide Field of View Achromatic Metalens

J. ENGELBERG,[1,2] R. MAZURSKI,[1] AND U. LEVY[1,*]

[1]*Department of Applied Physics, The Faculty of Science, The Hebrew University of Jerusalem, Jerusalem, Israel, 9190401*
[2]*Department of Electro-optics and Applied Physics, Jerusalem College of Technology, Jerusalem, Israel, 9116001*
*[\*ulevy@mail.huji.ac.il](mailto:ulevy@mail.huji.ac.il)*

**Abstract**

Metalenses have become ubiquitous in academic research and have begun to make their transition to industry. However, chromatic aberration still greatly limits the applications of metalenses. Achieving a wide field-of-view (FOV) is another challenge that has been dealt with successfully by using a removed stop, but when combined with broadband spectrum, lateral chromatic aberration severely limits performance. In this paper we tackle this grand challenge and present a comprehensive design methodology for a simultaneously wide-FOV and achromatic metalens which is inspired by the human visual system. As a design example, we present a metalens operating in the near infrared (NIR), with 10% relative spectral bandwidth (807-893nm), focal length of 5mm, F/5, and FOV of ±20˚. In particular, we show how to optimize the stop position, and correct the lateral chromatic aberration, both of which have not been reported in the past. In addition, we evaluate the performance of the metalens using accurate performance metrics, and demonstrate the improvement compared to a chromatic metalens. Our approach paves the way for the design of wide FOV metalenses that can operate over a relatively large bandwidth, effectively contributing to the widespread implementation of metalens science and technology.

# 1  Introduction

Metalenses have been attracting a lot of attention over the last decade, when they were implemented using dielectric materials thus obtaining high efficiency [1,2]. It was quickly realized that since metalenses are a type of diffractive lens (i.e., their phase function is not continuous but rather modulo a multiple of 2π), they are severely limited by chromatic aberration [3]. Thus, the race for implementing achromatic metalens began. Several methods were originally proposed: Spatial multiplexing [4,5], cascading [6] and dispersion engineering, where the latter quickly became the dominant method used [7–10]. At first it seemed that the problem was solved, but it was soon shown that dispersion engineering was limited to small metalens apertures [11], or more accurately, to low Fresnel numbers [12,13].

In addition to the work on correcting metalens axial chromatic aberration, a parallel path was being pursued to achieve wide FOV metalenses, necessary for most imaging applications. This entails correcting the monochromatic off-axis aberrations. A solution was found in [14,15] based on [16], where it was shown that a removed aperture stop located at the front focal plane of the metalens allows for correction of the third-order off-axis aberrations (coma and astigmatism). Of course, there is a price paid here: Two optical elements, spaced apart, are needed, rather than one (in the case of [14] there is another metalens located at the stop, but even the stop itself, as in [17], can be considered as another element). Thus, the metalens system is no longer so thin. On the other hand, even without the additional element there is still the distance between the metalens and the sensor, so the overall system length has only been approximately doubled. There have been attempts to remove this price-tag, by applying a synthetic removed aperture using total internal reflection [18] or wavevector filter [19]. The drawback of the first method is that it has large spherical aberration which degrades resolution severely. While this could be corrected with image processing using deconvolution, in a real scenario image quality would be limited by noise. The second method seems to work nicely but is limited to monochromatic applications.

The design of a metalens which is both wide-FOV and achromatic is very challenging. While the correction of chromatic aberration is challenging even when restricting its operation to on-axis applications, it becomes even more difficult off-axis. The reason for this is that the removed stop solution for the monochromatic off-axis aberrations introduces a new off-axis aberration: Lateral chromatic, i.e., the focal spot experiences a wavelength dependent lateral shift (on top of the axial shift, see Figure 1(b)). In principle, this aberration could be corrected using dispersion engineered nanostructures, as was used to correct the on-axis chromatic aberration. Yet, for a wide-FOV metalens with a removed stop, there is pupil wander at the metalens surface, increasing the size of the metalens (Figure 1). This, in turn, increases the effective Fresnel number of the metalens, making it more difficult to correct the chromatic aberration [20].

There have recently been some reports of wide-FOV achromatic metalenses, using a removed stop and dispersion engineering [21–24]. In this paper we describe an achromatic wide-FOV metalens for the near-infrared (NIR) spectral range, focal length 5mm, F-number 5 (F/5), and total FOV 40˚. The main novelty of this paper with respect to previous reports is that we account for lateral chromatic aberration and optimize the stop location. Additional contributions of this paper are: (a) We design the lens with large dimensions, appropriate for real-world applications, such as security cameras. (b) We evaluate the lens using real-world performance measures, that account not only for resolution, but also for signal-to-noise ratio (SNR) [25]. (c) We introduced a phase jump along the metalens aperture and found that we were able to significantly improve center-FOV performance using this method. This approach is inspired by nature, e.g. the human visual system, which provides high resolution in the center of the FOV (fovea region), at the expense of lower resolution in the peripheral region, which is mostly used for tasks such as orientation and motion detection.

## 2  Design – macro level

We designed a relatively large scale, 5mm effective focal length (EFL), F/5, 40° full-FOV, achromatic metalens. We chose a wavelength range centered on 850nm with a spectral width of 86nm (~10% relative spectral width), which can be relevant for day or night-time surveillance (the latter with an appropriate band-pass filter and LED illumination). The most successful method proposed to date to correct monochromatic off-axis aberrations and obtain a wide-FOV using flat lenses is the landscape telecentric lens originally proposed by Buralli and Morris [16]. This method was implemented for a metalens doublet in [14] and for a single metalens in [15]. The method consists of an iris located at the front focal plane of the metalens, as shown schematically in Figure 1. Note the pupil wander, i.e., the location of the off-axis beam on the lens aperture shifts away from the axis, increasing the required metalens diameter.

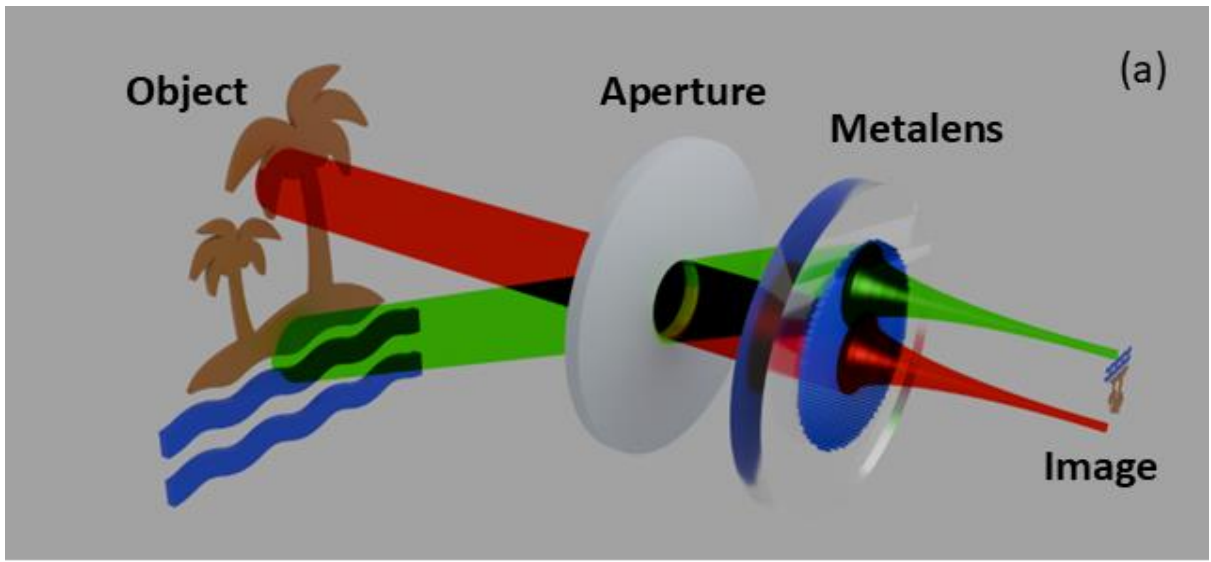

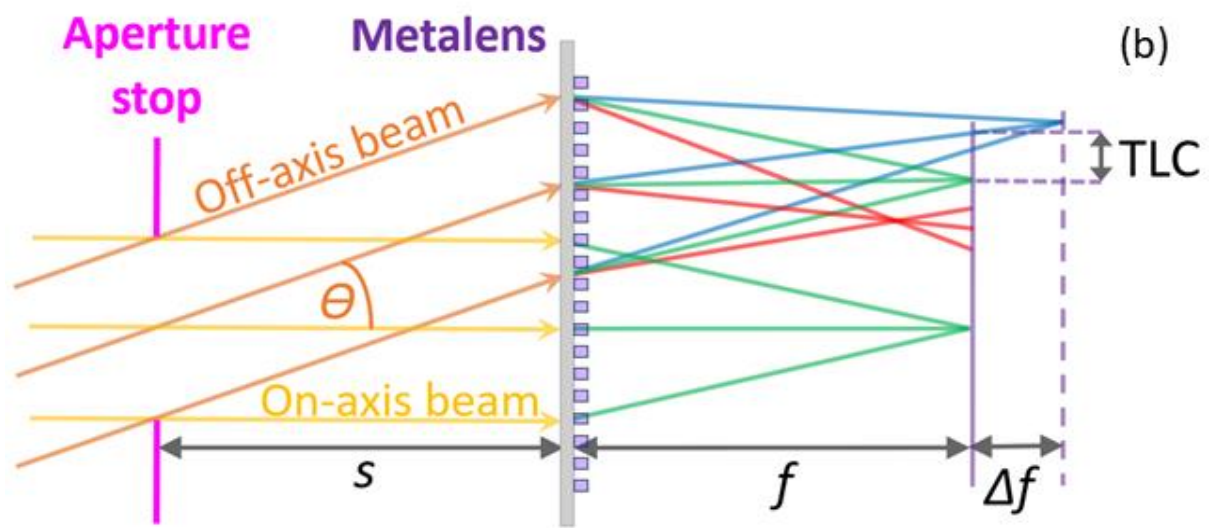

*Figure 1: (a) Sketch of landscape type metalens system. (b) Drawing showing off-axis chromatic aberrations at the center wavelength focal plane, located at a distance f to the right of the lens, for the case of a removed aperture stop located at distance s to the left of the lens. Both axial (defocus Δf of blue and red rays) and lateral (displacement of chief ray, TLC, for the extreme wavelengths along the radial direction) chromatic aberrations affect the off-axis beam.*

The landscape lens method has two main drawbacks: (a) The larger the FOV, the larger the pupil wander on the metalens surface, requiring a large metalens aperture. (b) Lateral chromatic aberration, proportional to the stop distance *s*, is introduced off-axis as shown in Figure 1(b) [3,15]. These drawbacks can be mitigated if we can place the stop nearer to the metalens, with the additional advantage of decreasing overall system length. The second drawback can be additionally diminished if we are able to somehow achromatize the metalens. The second drawback indicates that the optimal stop location for a broadband metalens may be closer to the metalens than for a monochromatic metalens, since there is a tradeoff between the monochromatic and chromatic aberrations as we shift the stop position [3].

Buralli and Morris referred to the case of a monochromatic metalens with quadratic phase ($\phi(r) = Ar^2$) and concluded that the optimal stop position is at the front focal plane of the diffractive lens, resulting in perfect correction of all the off-axis third order aberrations (except distortion, but this aberration does not affect resolution). However, by using higher order phase coefficients ($\phi(r) = A_1 r^2 + A_2 r^4 + \cdots$) the correction of the monochromatic aberrations at closer stop distances can be significantly improved. We demonstrate this capability in Figure 2, which compares the modulation transfer-function (MTF) for a quadratic vs. high-order polynomial phase metalens. The design and simulation were done for monochromatic illumination (850nm), using Zemax optical design software.

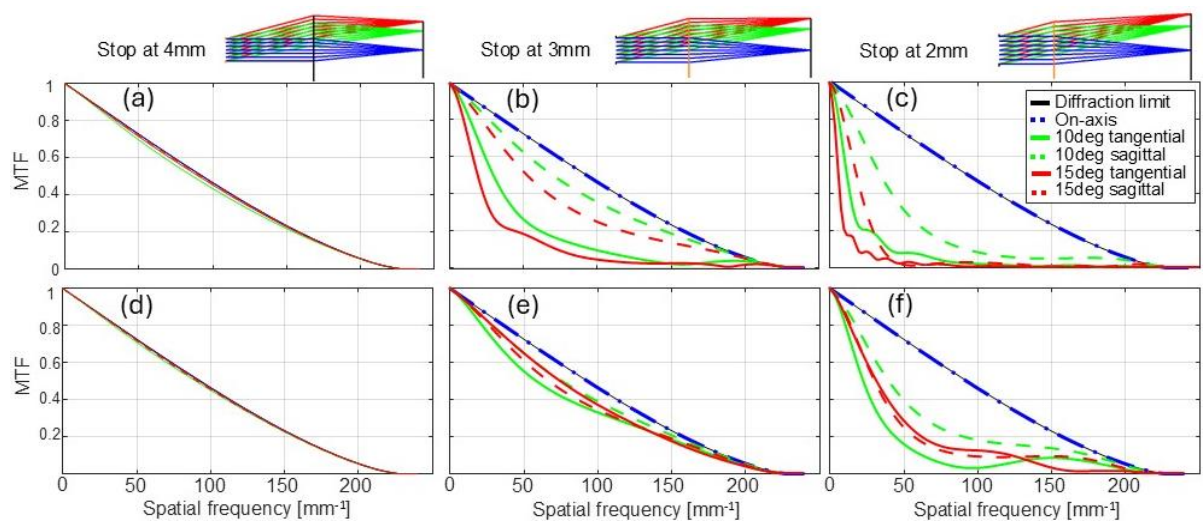

Figure 2: Monochromatic MTF at different stop distances for our 5mm EFL F/5 metalens – comparison of quadratic to high-order phase function. (a-c) MTF for a quadratic phase, at stop distances of 4, 3 and 2mm from the metalens respectively. (d-f) MTF for optimized high-order phase, for the same stop distances. The insets at the top show the design layout for the different stop distances.

For a monochromatic metalens, the optimal stop position from an optical performance point-of-view is clearly still at the front focal plane. However, for the sake of compactness, it may well be desired to use the design of Figure 2(e). As can be seen, the degradation of optical performance is negligible, whereas the overall length of the optical system, from stop to image, can be reduced by about 20%, from ~10mm to ~8mm. Note that in our design we neglected the substrate thickness, taking it as zero. Neglecting the substrate thickness allowed us to better demonstrate the dependence of performance on stop position, without the extra complication of optical thickness of the substrate. In real metalens we will of course need to introduce a mechanically feasible thickness. In Fig. S1 and S2 we present the optical prescription of the design shown in Figure 2(e), with a stop distance of 3mm, and an equivalent design with a 1mm sapphire substrate and the stop 2.5mm away from the substrate front surface.

So far, we have discussed monochromatic operation of our metalens. For a polychromatic metalens, the optimal stop position will likely be obtained at a shorter stop distance, as this decrease in stop distance allows reduction of the lateral chromatic aberration. To demonstrate this, we first look at the extreme case of a 'chromatic' metalens, meaning a metalens with no chromatic correction. In this case the optimal stop position will be closest to the metalens, to reduce the lateral chromatic aberration. Zemax simulation results are shown in Figure 3, for 21 equally spaced wavelengths in the range 810-890nm. It turns out that the optimal stop position for chromatic metalens, based on the Strehl ratio at 15˚ FOV, is only 2mm away from the lens. While this is difficult to discern visually from the point-spread functions (PSFs) and MTFs shown in Figure 3, it can be easily extracted from the numerical Strehl ratio values in Figure 9(b). Note that while the variation in performance around the optimum stop location is fairly small, one can still observe (Figure 3(a-c)) that the performance goes from being limited by lateral chromatic at 3mm distance (more elongated PSF) to being limited by coma at 1mm distance (more comatic shape).

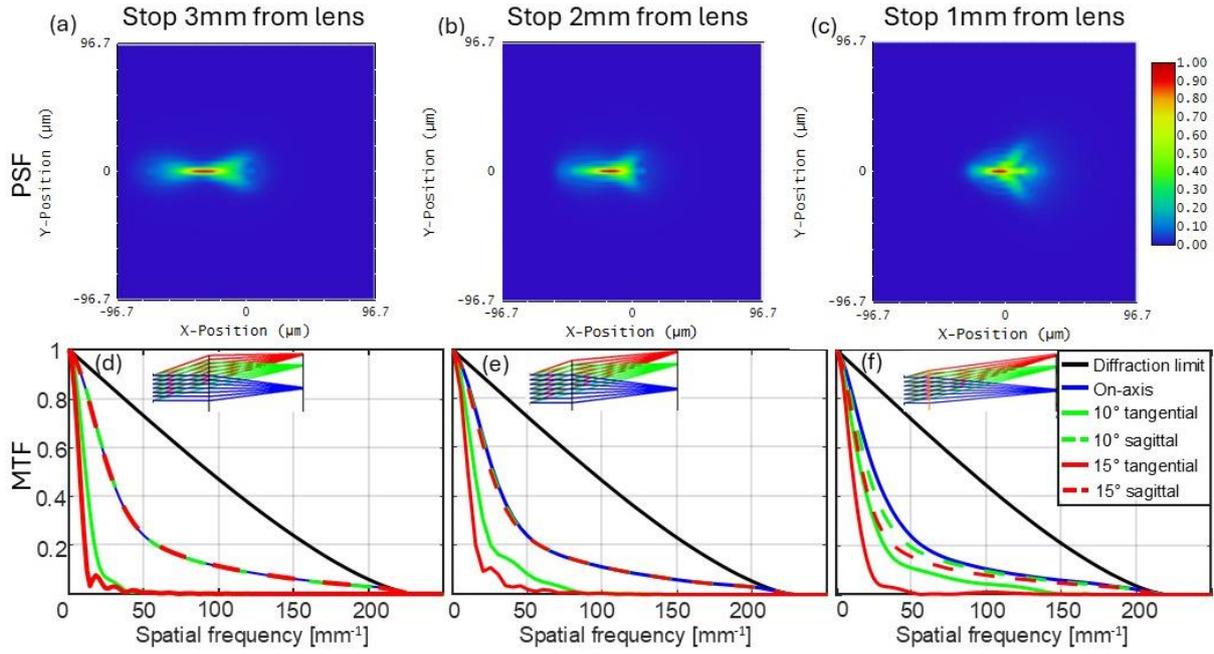

*Figure 3: Polychromatic (810-890nm) PSFs and MTFs at 15° field point for different stop locations. (a,d) PSF and MTF for stop distance of 3mm respectively. (b,e) PSF and MTF for stop distance of 2mm respectively. (c,f) PSF and MTF for stop distance of 1mm respectively. The optimum Strehl ratio is obtained at stop distance of 2mm. The insets show the design layout for the different stop distances.*

## 3 Design – nano level

Following the macro-level metalens design, we now discuss the nano-level design, allowing for chromatic correction, based on dispersion engineering [8,9,11]. Normally, when light propagates a distance *H* through a truncated waveguide type nanostructure the phase dispersion ($d\phi/d\omega$) is positive, because $\phi(\omega) = \frac{\omega}{c} n_{eff}(\omega) H$ and $dn_{eff}/d\omega$ is positive (the larger the frequency the higher the mode confinement, so the higher the effective refractive index, $n_{eff}$) [13]. This type of dispersion is useful for correcting the chromatic aberration of a diffractive lens, since the higher frequencies (shorter wavelengths) tend to focus less (weaker diffraction), so we need to give them more phase. However, as was shown by Shrestha et el. [11], if the zero-phase point is defined at the center of the lens, the phase induced by a positive (focusing) metalens needs to be negative. This means that to correct chromatic aberration we need the phase to be more negative (larger absolute value) for higher frequencies, i.e., we require *negative* phase dispersion. Since such structures are not available, it was suggested by Shrestha to overcome this by adding a frequency dependent phase bias, $C(\omega)$, which shifts the zero phase to the edge of the lens. Now the required dispersion is positive, which can be accommodated by typical nanostructures. We used this method, of applying different phase pistons for the different wavelengths, so that the phase for all wavelengths is identical at the edge of the aperture.

While the above achromatization method can work well for small scale lenses, when we try to design a large diameter metalens, the rapidly changing phase for each wavelength as we move away from the optical axis, necessitates high-dispersion nanostructures near the center of the lens so that each wavelength can follow its phase function. If the required dispersion exceeds the theoretical limits [12], or those of the specific nanostructure library, the metalens performance will degrade. Therefore, as the diameter of the metalens increases, it becomes more difficult to obtain good polychromatic performance.

As can be seen in Figure 1, despite our modest F-number of 5, the FOV requirement increases the metalens diameter because of pupil wander, making it difficult to achieve good achromatic performance. We found that when we increase the diameter of the metalens to obtain large FOV, we damage the on-axis imaging, as it becomes harder to fit the correct phase for all the wavelengths simultaneously. However, we can leverage the fact that each field-point only uses part of the aperture, to introduce a phase jump at a certain aperture radius. The phase jump is implemented by changing the phase bias *C(ω)* at this aperture radius. The jump will only affect intermediate field points that use a section of the aperture that includes the said aperture height radius. We selected the jump location to be at the edge of the on-axis beam, thus obtaining characteristics which are like the human vision system.

This method allows us to improve the overall performance, with the greatest improvement occurring on-axis, whereas the off-axis resolution is of lesser quality, and is mostly used for secondary tasks such as orientation and motion detection. More detailed information about our nano scale design can be found in the Materials and Methods section.

## 4  Results

The performance of our achromatic metalens is evaluated based on the metrics introduced in [25]. These metrics include Strehl ratio as a measure of resolution, diffraction efficiency ($\eta$) and transmission ($T$) as a measure of signal-to-noise ratio (SNR), and the overall-performance-metric (OPM) that combines them into a single performance metric, according to Eq. (1):

$$OPM = \frac{\eta}{\sqrt{T}} \cdot Strehl \tag{1}$$

The performance was evaluated at 33 wavelengths within our spectral range (807-893nm), and their results were averaged to obtain the polychromatic performance. The PSF for each wavelength was calculated using angular spectrum propagation [26] from the metalens to the image plane, and the PSFs were summed to obtain the polychromatic PSF. The pupil phase function after the metalens for each field point was calculated based on the geometrical path length of the rays accounting for the wavefront tilt of each field point (multiplied by $2\pi/\lambda$ to convert to the path length to phase for each wavelength $\lambda$), and adding the metalens phase at that wavelength, for each radial position on the metalens, as described by Eq. (3) in the Materials and Methods section.

The overall transmission $T$ is calculated based on the spectral transmission of each of the nanostructures calculated in Lumerical Finite-Difference-Time-Domain (FDTD) software. The diffraction efficiency $\eta$ is the fraction of light arriving at the first order of diffraction based on the angular propagation simulation result, relative to the incident light. To calculate it, we first calculated the 'focusing efficiency', which we define as the fraction of light arriving at the first order of diffraction relative to the light transmitted through the metalens. Denoting the focusing efficiency as $\eta_f$, one obtains: $\eta = \eta_f T$. The focusing efficiency is calculated as the sum of the PSF values, within a small region in the image plane, until the PSF intensity flattens out [27], divided by the sum of the entire simulation region on the image plane (a square with side-length of 4mm which is 4 times the aperture diameter, so it includes all the zero and second order light, and much of the other orders). Note that this definition of focusing efficiency is different from that commonly used by other authors, which represents an unknown mixture of resolution and efficiency, and we do not recommend using [27]. The Strehl ratio used here is the volume under the 2-dimensional (2D) MTF relative to the volume under the diffraction limited MTF (the official definition is based on PSF peak, but the MTF based definition is equivalent for the on-axis case, and more suitable for our purposes in the off-axis case [25]).

The PSF of our achromatic metalens on-axis is shown in Figure 4, compared to the PSF of the equivalent chromatic metalens. The achromatic metalens shows a similar full-width at half-maximum (FWHM) value, but reduced energy in the sidelobes, which corresponds to the higher PSF peak (Figure 4,) and higher polychromatic Strehl ratio, as shown later in Figure 6 This is a result of correction of the axial chromatic aberration. All PSFs shown in the main manuscript are polychromatic. PSFs for three discrete wavelengths (center and extreme) are shown in Figure S9.

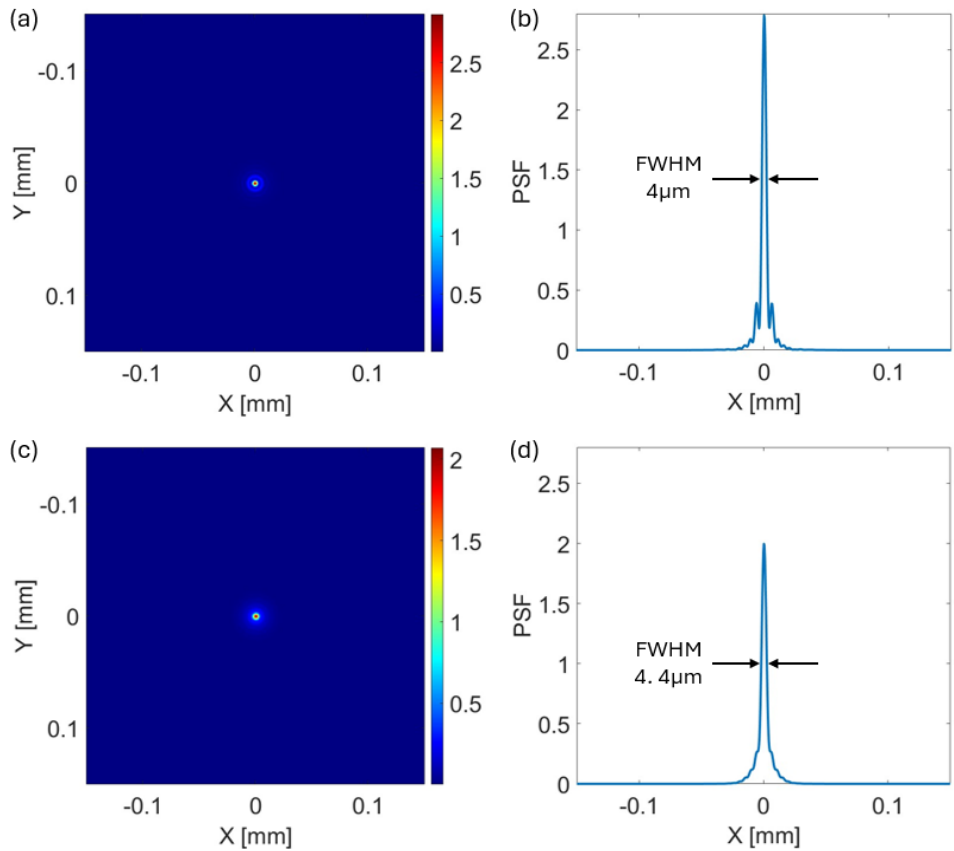

*Figure 4: On-axis PSF comparison between chromatic and achromatic metalens. (a-b) Achromatic metalens 2D PSF and cross-section respectively. (c-d) Chromatic metalens 2D PSF and cross-section respectively. FWHM is marked on PSF cross sections. All PSFs are normalized so that the volume under the 2D PSF function is equal to 1.*

Figure 5 compares the metalens MTFs. MTFs at three specific wavelengths (center and extreme wavelengths) are shown. In addition, the polychromatic MTF, averaged over 33 wavelengths within the range with uniform weighting, is shown. The chromatic metalens (panel b) has diffraction-limited (DL) MTF at the central wavelength (850nm), but poor MTF at the extreme wavelengths because of chromatic aberration, resulting in poor polychromatic MTF. In contrast, while the achromatic metalens MTFs are below DL, they are similar for all wavelengths (panel a), resulting in an improved polychromatic MTF (27.4% vs. 22.8% at 50c/mm, about the same at 100c/mm, but again much improved at 150c/mm and diffraction limited above 190c/mm).

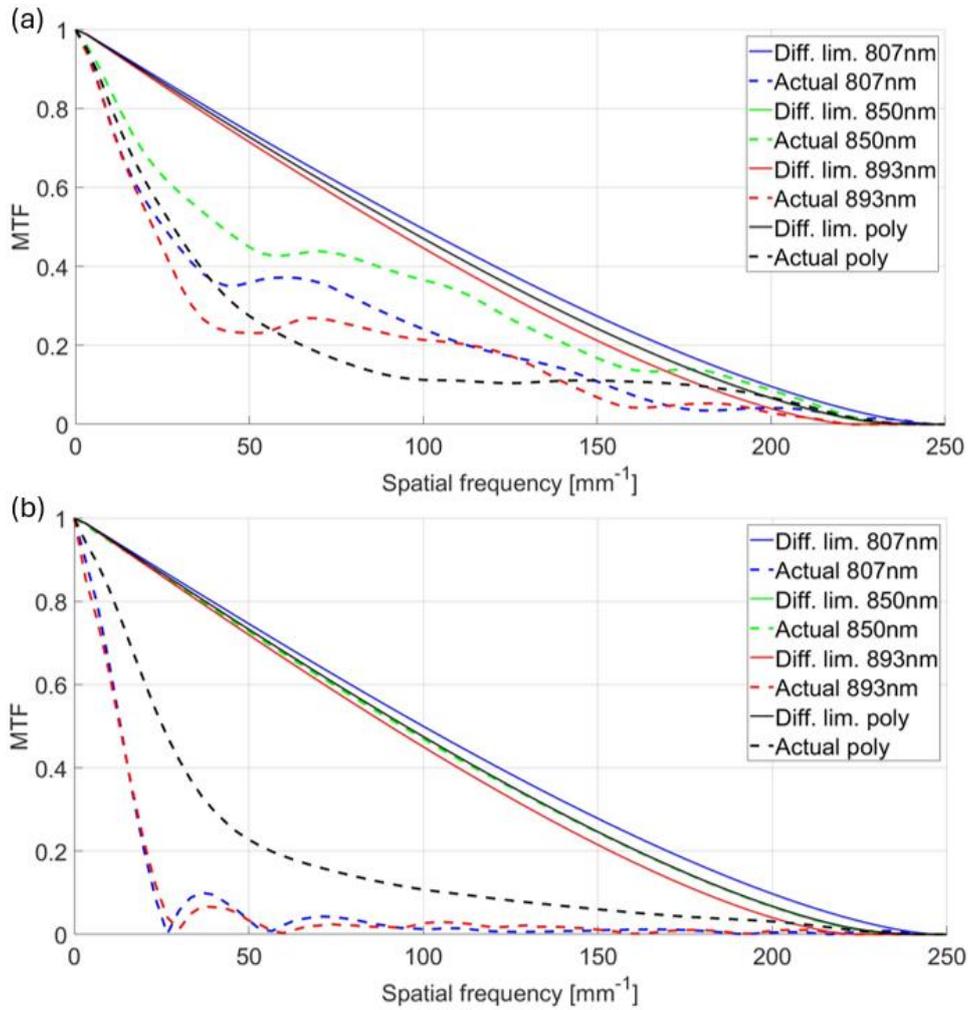

*Figure 5: On-axis MTF comparison. (a) Achromatic metalens. (b) Chromatic metalens. Solid lines represent diffraction limited MTFs, and dashed lines are the achieved MTFs.*

Figure 6 compares the Strehl ratio, transmission, focusing and diffraction efficiency, and OPM as a function of wavelength for achromatic (panel a) vs. chromatic (panel b) metalenses. In addition to the results for each wavelength we also show the polychromatic results for these metrics, which are represented as a flat line, as they are not wavelength dependent, but are rather the average over all the wavelengths. Here too, the chromatic metalens achieves excellent results at the nominal wavelength but degrades quickly as the wavelength departs from the nominal, so that the polychromatic performance is poor. In contrast, our achromatic metalens gives more uniform performance over all wavelengths, achieving a polychromatic OPM of 0.33, i.e. two-fold improvement compared to 0.17 for the chromatic metalens.

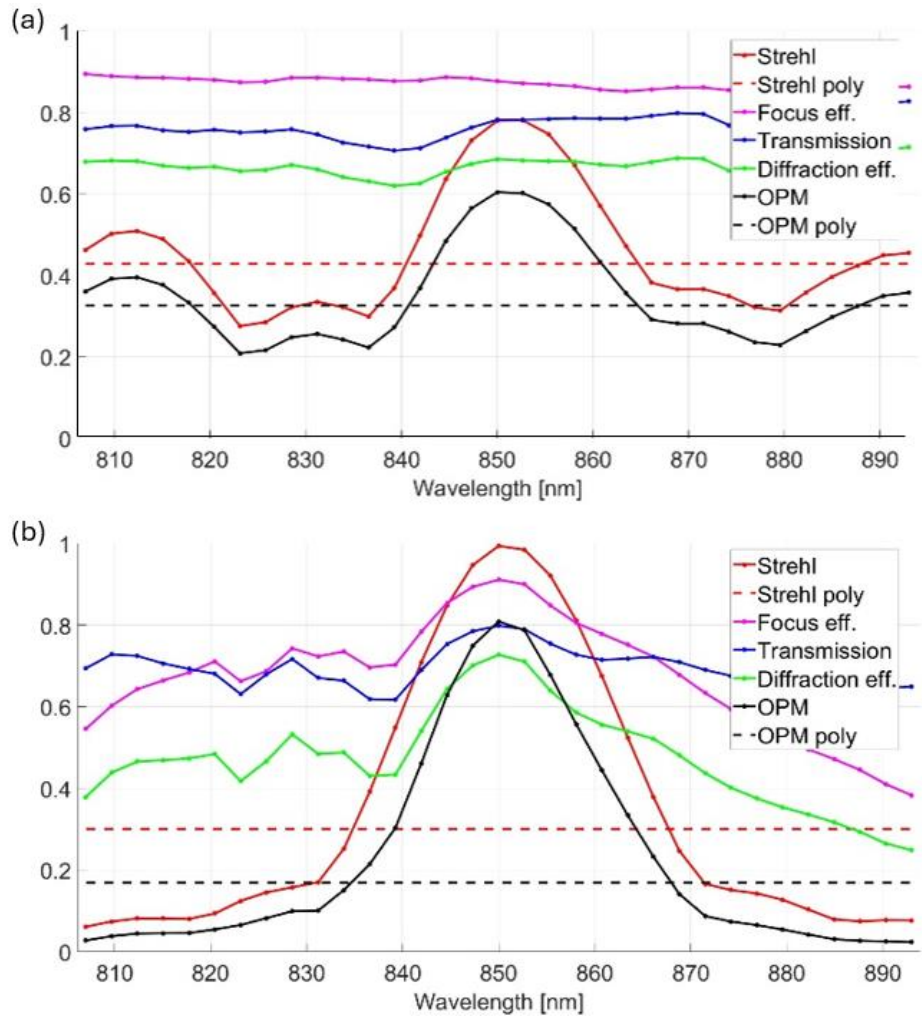

*Figure 6: On-axis Strehl, efficiency and overall performance (OPM) comparison as a function of wavelength. (a) Achromatic metalens (b) Chromatic metalens.*

Off-axis the performance is dependent on the stop distance from the lens, as shown in Figure 3 for the case of a chromatic lens. The optimal stop position is the one where the lateral chromatic aberration (which is smaller for short stop distance) and the monochromatic off-axis aberrations (which are smaller for long stop distance) are balanced. For an achromatic metalens we can expect the optimal stop distance to be farther from the metalens than for the chromatic metalens, since the lateral chromatic aberration is reduced.

Figure 7 and Figure 8 show results for off-axis performance, at 15° field angle, using the same performance parameters as for the on-axis case. The results shown are for the optimal stop for both metalenses, which turns out to be 2.5mm for the achromatic metalens and 2mm for the chromatic metalens - see Figure 9(b). For the off-axis case, the average Strehl/OPM is significantly worse than would be expected by looking at the average of the values for each wavelength. This is because the lateral chromatic aberration introduces a lateral shift between the PSFs of the different wavelengths, meaning that despite good Strehl per wavelength, their spots don't overlap, so their sum gives a large blur in the radial direction. The MTFs shown are only in the radial (tangential) direction, but the Strehl ratio is calculated based on the 2D MTF, so it and the OPM are accurate. Polychromatic performance for other field angles, of 5°, 10° and 20°, is shown in Figures S3-S8 respectively. Off-axis performance at single wavelengths, for field angles of 5°, 10°, 15° and 20°, is shown in Figures S10-S13 respectively.

An interesting point to note is that despite the modest improvement in polychromatic Strehl ratio, there is a significant change in the shape of the PSF cross section in the radial direction. It transforms from a more rounded peak in the chromatic case to a sharp peak in the achromatic case. This leads to a significant decrease in the FWHM of the PSF cross-section, as shown in Figure 7 and Figure 11. While in [25,27] it was explained that in general the Strehl ratio is a better resolution performance metric than FWHM, there may be applications where the FWHM is important (for example if we can threshold the image, ignoring whatever is below a given intensity, as is often done in imaging-based lithography systems).

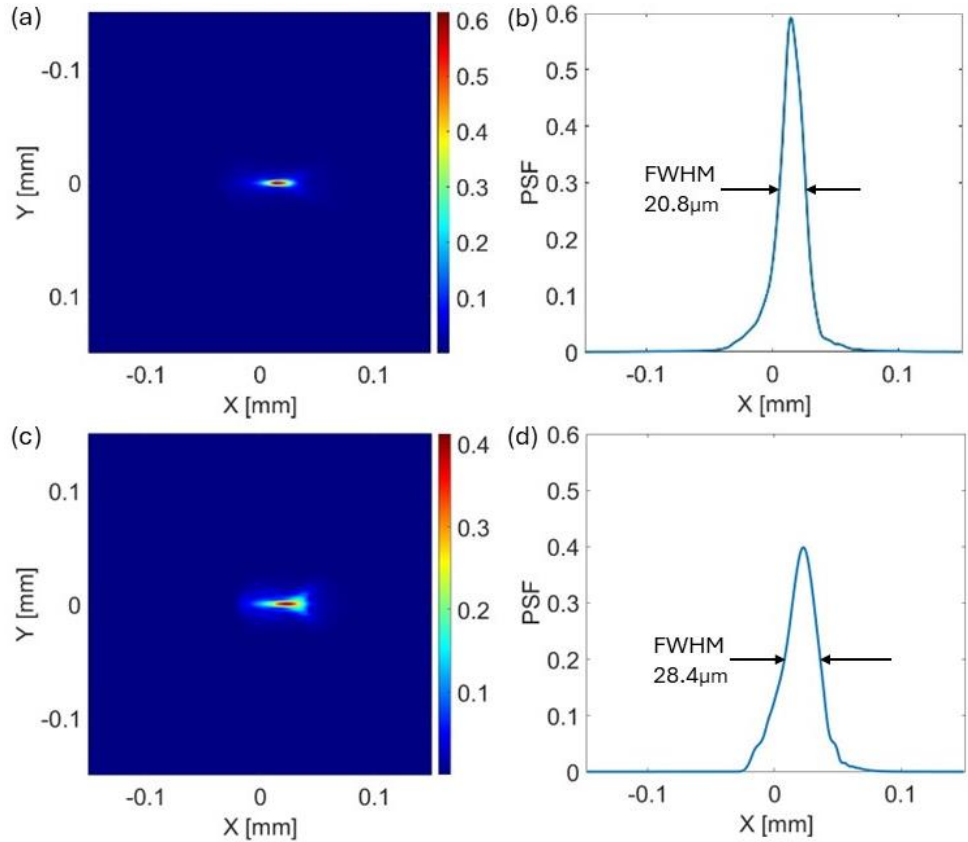

*Figure 7: 15˚ off-axis PSF comparison. (a,b) Achromatic metalens 2D PSF at the optimal stop distance of 2.5mm and cross-section in the radial (horizontal) direction respectively. (c,d) Chromatic metalens 2D PSF at the optimal stop distance of 2mm and cross-section in the radial (horizontal) direction respectively.*

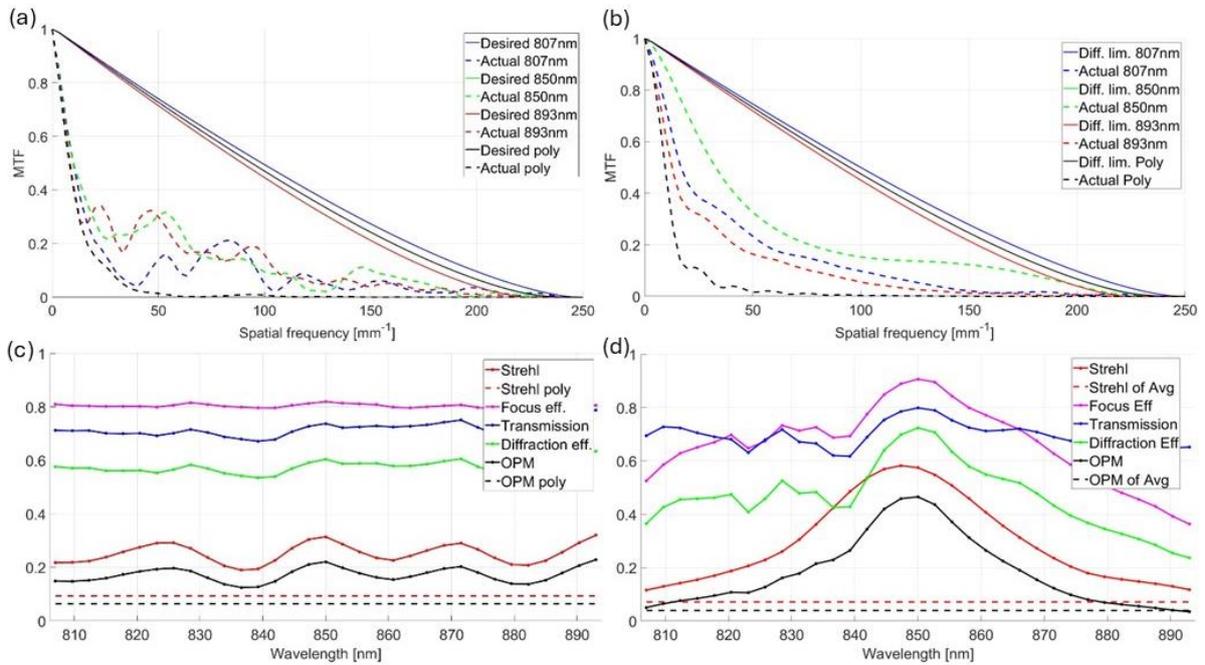

*Figure 8: 15˚ off-axis performance comparison at optimal stop distance for each metalens (2.5mm for achromatic and 2mm for chromatic). (a) Achromatic metalens MTF (b) Chromatic metalens MTF (c) Achromatic metalens spectral and polychromatic Strehl, efficiency and overall performance (OPM) results as a function of wavelength (d) The same for chromatic metalens.*

Figures 9-12 compare the performance of the achromatic vs. chromatic metalens in terms of Strehl ratio, diffraction efficiency, FWHM and OPM, respectively, for various stop distances and field angles. Figure 9(a) shows that for all cases the achromatic metalens gives better resolution than its chromatic counterpart, yet the improvement is moderate. Figure 9(b) shows the cross-section of 12(a) at FOV of 15°, which we chose as our reference field-point for determining optimal stop position.

As opposed to the moderate improvement in Strehl ratio, the improvement in diffraction efficiency, shown in Figure 10, is dramatic. This is of crucial importance, in particular for low light applications, where every photon counts. Furthermore, while for small field angles the improved Strehl of the achromatic metalens does not significantly impact the radial FWHM, shown in Figure 11, for large field angles the relative improvement in FWHM is significantly larger because of the lateral chromatic aberration correction. As previously mentioned, this might play a major role for applications where thresholding is applied. Finally, the overall performance and its relative improvement are shown in Figure 12, demonstrating a major improvement over the chromatic metalens.

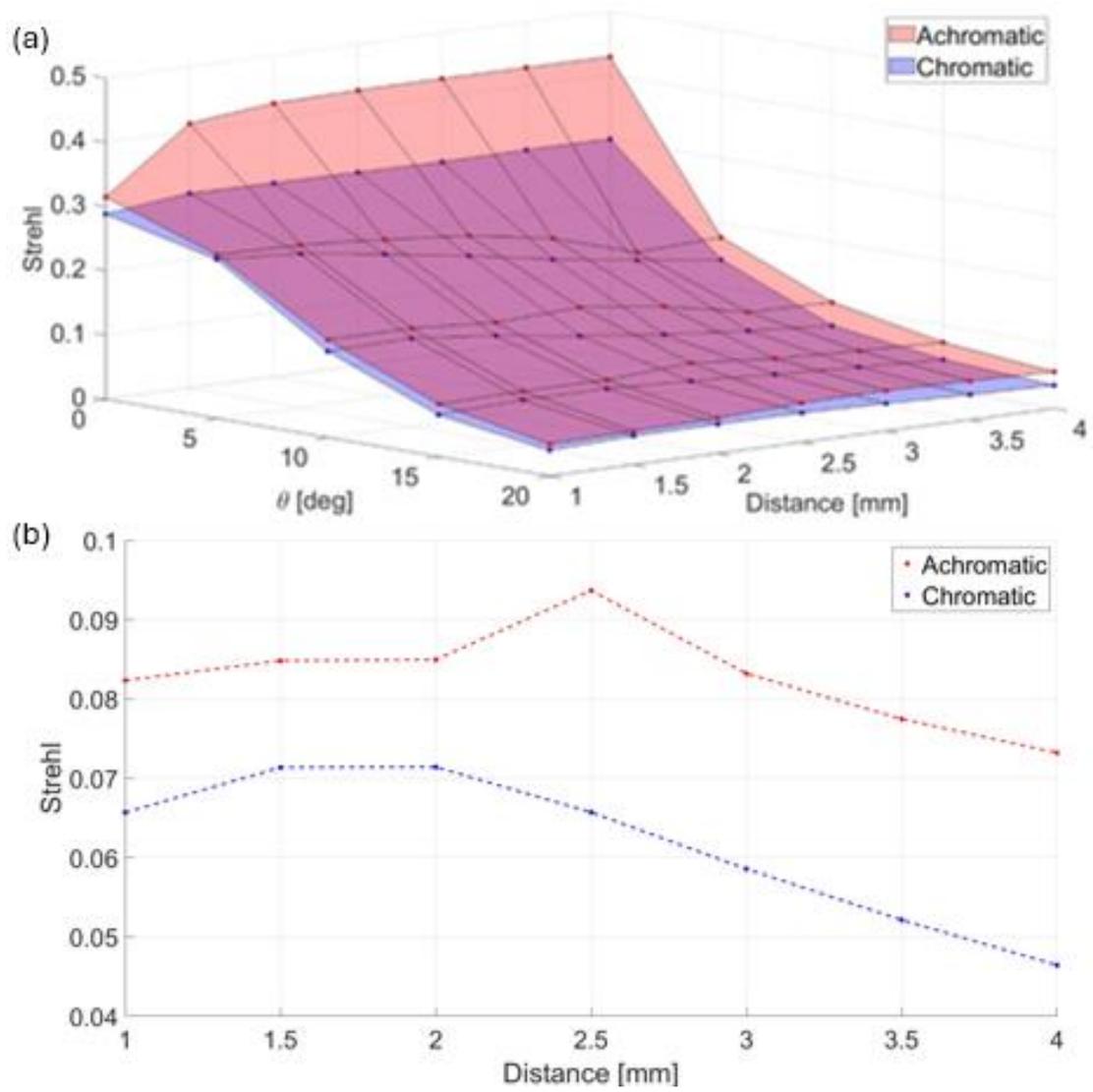

*Figure 9: (a) Strehl ratio as a function of field angle (Θ) and stop distance. (b) Cross-section of (a) at 15 degrees.*

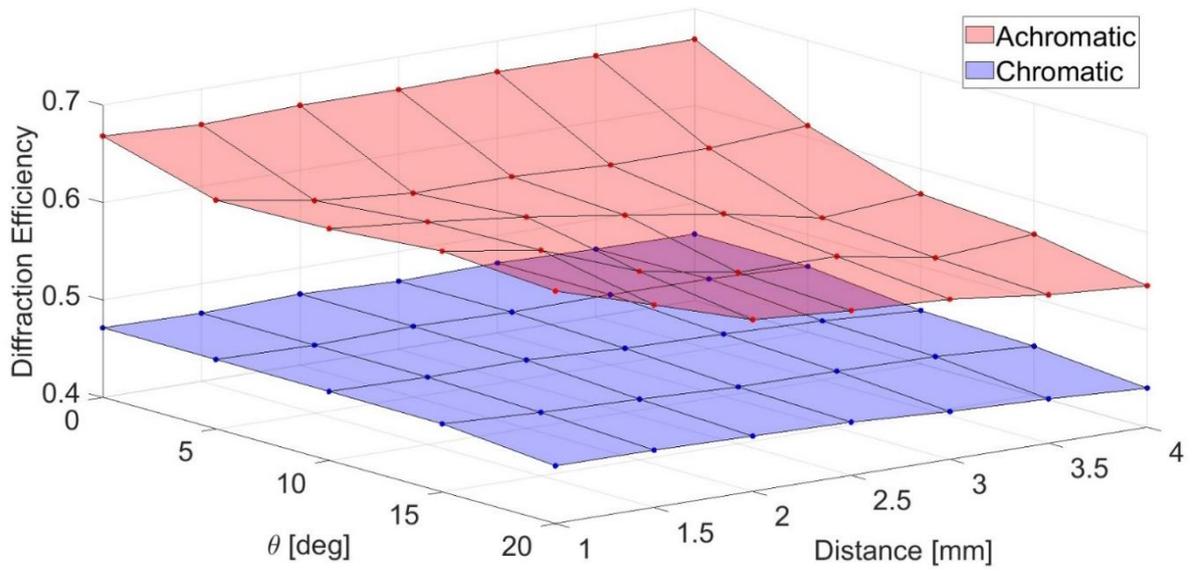

*Figure 10: Diffraction efficiency as a function of field angle (Θ) and stop distance.*

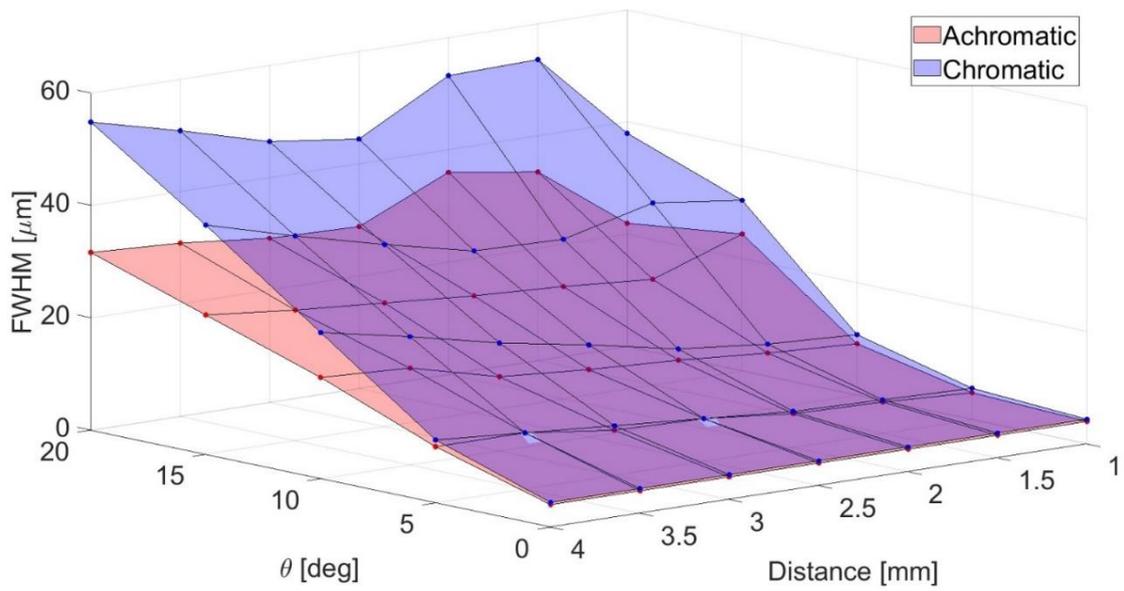

*Figure 11: FWHM in the radial direction as a function of field angle (Θ) and stop distance.*

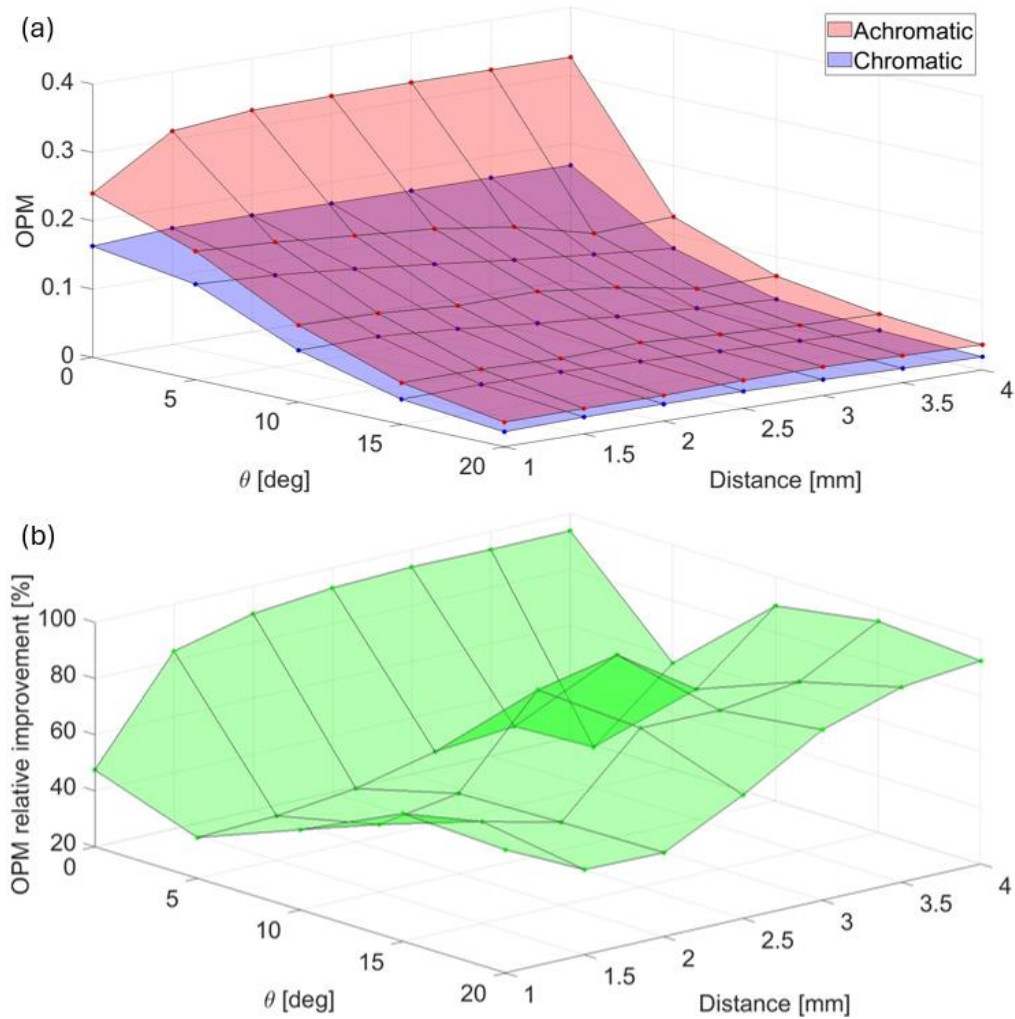

*Figure 12: (a) Overall performance as a function of field angle (Θ) and stop distance. (b) Relative improvement of OPM as a function of field angle (Θ) and stop distance.*

# 5 Discussion

The achromatic metalens presented in this paper was designed with parameters relevant to real world applications, such as a security camera lens. The emphasis of the paper is less on the specific performance achieved, but rather on the design methodology (e.g. use of a phase jump, optimizing the stop position, use of polychromatic PSF that includes the effect of lateral chromatic aberration, improving on axis resolution, comparison to chromatic baseline), use of good performance metrics, and understanding the tradeoffs.

While the importance of a removed stop has long been understood [3,14–16], it was generally accepted that it should be located at the front focal plane of the metalens to cancel the monochromatic third-order aberrations of coma and astigmatism. While there have been some publications where it was located closer to the metalens [22–24], no explanation was offered as to why this location was chosen. Here we explained the tradeoffs for the case of a narrowband metalens (compactness vs. performance) and for a broadband metalens (lateral chromatic aberration vs. monochromatic aberrations) and demonstrated an optimization method.

As explained in section 3, we used a single phase-jump in our metalens, at an aperture radius equal to that of the on-axis beam. Phase jumps have been implemented in some previously reported metalenses to compensate for the nanostructure phase dispersion, compared to required phase dispersion. In [28] this was done for the case of narrow-FOV metalens operating at 3 discrete wavelengths. The phase jump location and magnitude (for each wavelength) were optimized to obtain optimum performance. In [24] phase jumps were used for a wide-FOV metalens doublet operating in the NIR, with phase jump locations determined by folding the ideal phase dispersion onto the available phase dispersion of the nanostructure library. In contrast, we use a nature inspired approach, providing optimal resolution in the center of the FOV by using a single-phase jump. By sacrificing resolution in the peripheral FOV we gain high resolution in the center of the FOV, and much higher efficiency over the entire FOV (about 60% in our design as compared to about 15% in previous designs).

Our results demonstrate that for most real-word applications, which require a focal length in the order of at least a few millimeters and a significant FOV, chromatic aberration is still a severely limiting problem. This is despite the plethora of reports on achromatic metalenses.

It would be interesting to compare our performance to that of other published broadband wide-FOV metalenses. Each publication has different parameters, so comparing the OPM directly would not be useful. However, in a previous publication we introduced an extended OPM metric (EOPM) that accounts for the scaling of the metalens (via the focal length) and the relative spectral range [25]. What prevents proper comparison between different published designs is that most often proper performance metrics are not used [27]. For characterization of resolution, FWHM of the PSF is commonly used, instead of the more appropriate Strehl ratio. Instead of diffraction efficiency, 'focusing efficiency' is usually used, which is defined as the percent of energy in a specific spot radius, without regard to whether the PSF has flattened out by that radius or not. This leads to a metric that mixes diffraction efficiency with resolution, which cannot be used in Eq. (1). In addition, the overall transmission is usually not reported.

## 6   Materials and Methods

This section gives details of our nanostructure library design. To correct the chromatic aberrations of our metalens, we designed and simulated a library of periodic silicon nanostructures on a sapphire substrate using Lumerical Finite-Difference-Time-Domain (FDTD) software. The nanostructures were designed as truncated waveguides with varying shapes, as shown in Figure 13. The period (P) and the height (H) are constant for all pillars and are set to 0.4μm and 1.5μm respectively. For the circular pillars, the radii dimensions were swept in the ranges: $0 < R_3 < 160nm$, $R_3 < R_2 < 180nm$, $R_2 < R_1 < 200nm$. For the square pillars, the widths were swept in the ranges: $0 < D_3 < 320nm$, $D_3 < D_2 < 360nm$, $D_2 < D_1 < 400nm$. The dimensions $R_i$, $D_i$ are as shown in Figure 13(g-l).

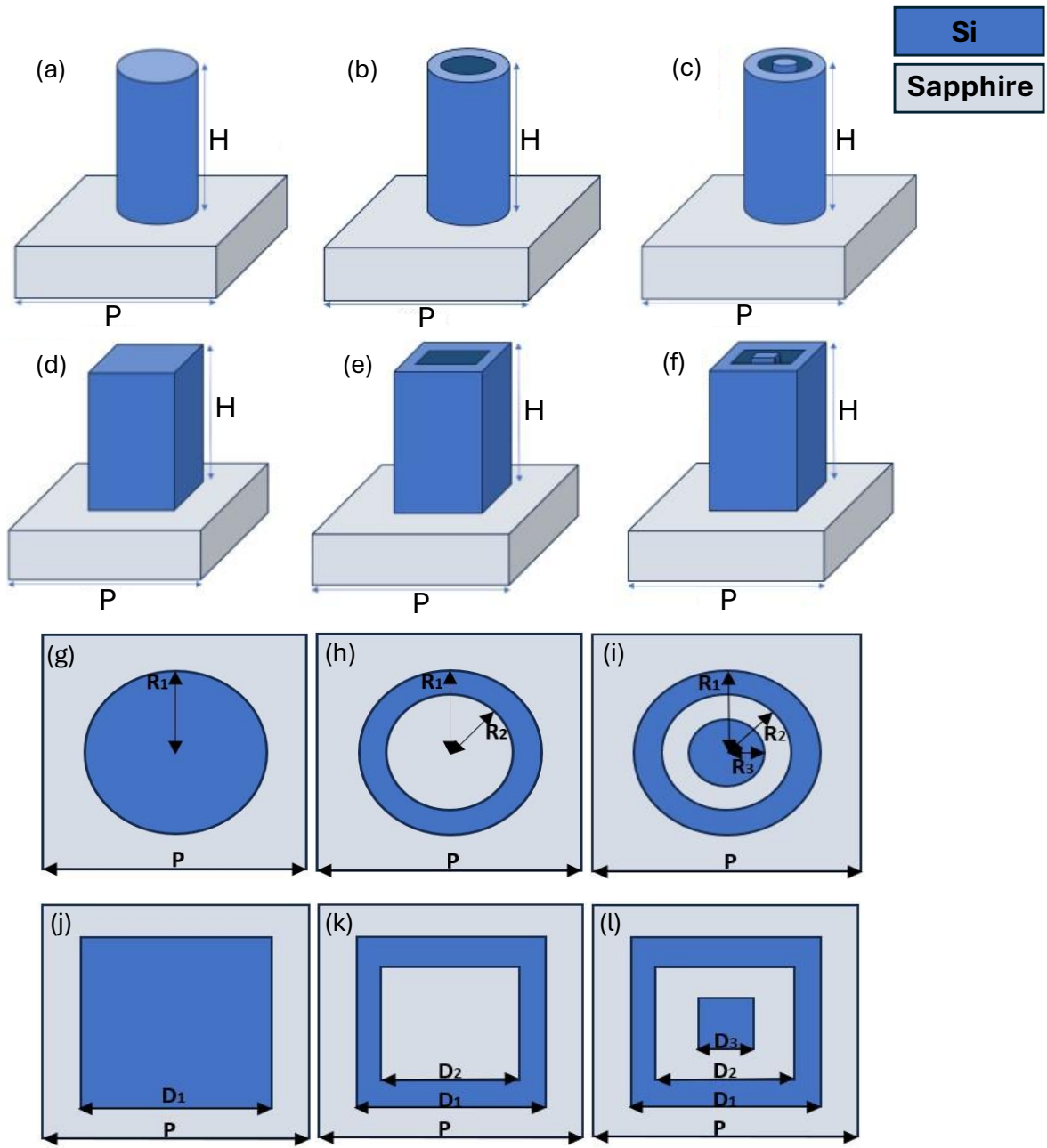

*Figure 13: Nanostructure library for achromatic metalens. (a-f) Perspective views of the nanostructures. (g-l) Top views of the same nanostructures.*

To evaluate our nanostructure library, we created the graph presented in Figure 14(a), that shows the dispersion, Δn, as a function of the phase, ϕ, of the center wavelength. The dispersion is represented by the difference in effective refractive index between the extreme wavelengths. Each circle on the graph represents a specific nanostructure geometry. We display and use only nanostructures that provide transmission of above 40%. In the following we compare our achromatic metalens design to that of an equivalent chromatic metalens. For the chromatic metalens we used simple cylindrical pillars, as presented in tiles (a) and (g) of Figure 13. Figure 14(b) shows the sparse dispersion graph for this library. The nanostructure phases at the various wavelengths were evaluated using Lumerical FDTD software, at normal incidence only, under the assumption that the phase does not change much for other incidence angles. This allowed us to simplify the analysis, and is reasonable, since it was shown in [29] that at moderate incidence angles (up to 20˚) and NA (up to 0.13) metalens performance is retained, indicating that the phase is close to that of normal incidence.

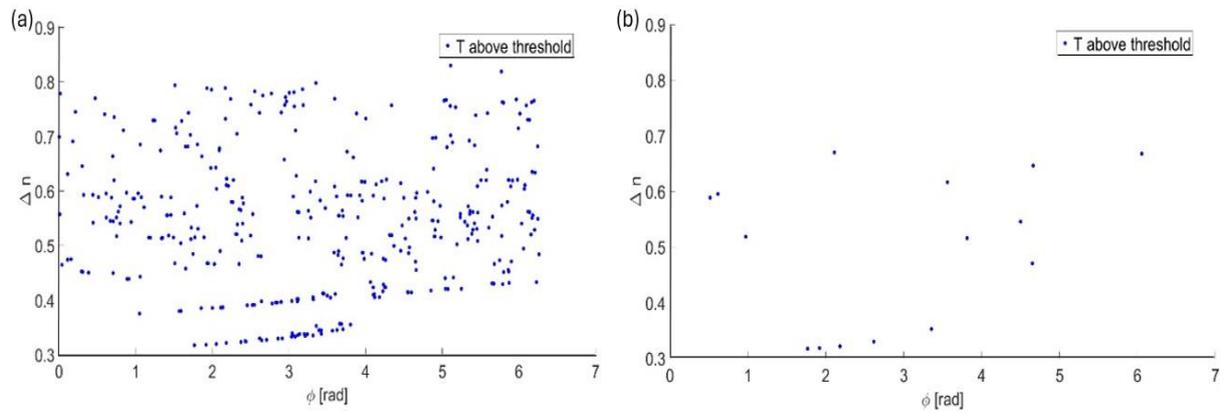

Figure 14: Dispersion in terms of Δn as a function of ϕ for (a) achromatic nanostructure library, and (b) chromatic nanostructure library.

The achromatic metalens bandwidth limit formulated in [12] was translated to a limit on the maximum attainable Fresnel number (FN) of a diffraction limited achromatic metalens in [13], and is expressed in Eq. (2):

$$FN_{max} = 2h\Delta n \left( \frac{\Delta \lambda}{\lambda} \right)^{-1} \qquad (2)$$

Where $h$ is the pillar height relative to the wavelength, $\Delta n$ is the maximum dispersion, and $\Delta\lambda/\lambda$ is the relative spectral range. For our lens parameters we have $h = \frac{H}{\lambda} = \frac{1500}{850} = 1.76$, and $\Delta\lambda/\lambda=0.1$. For $\Delta n$ the theoretically obtainable dispersions are in the range of 0 to 2.1. This is because our nanostructures are made of Silicon (refractive index @850nm ~3.6) and in our design we used a cover of PMMA (refractive index @850nm ~1.5). However, in practice the dispersion will always be lower because of fabrication constraints, such that the dimensions of the radii and the difference between them cannot approach zero. And indeed, as can be seen in Figure 14(a) we obtain a maximum dispersion of 0.83, and a dispersion range of 0.51 (since the minimum dispersion is not zero, but 0.32). Based on the above formula, and using Δn=0.51, the maximum FN comes out to be 18.

Let us compare this to the FN of our lens, which is given by [26]: $FN = \frac{R^2}{\lambda f}$, where $R$ is the lens aperture radius. In our case we have $R = \frac{D}{2} = \frac{f}{2F} = \frac{5}{2 \cdot 5} = 0.5mm$, and $FN = \frac{0.5^2}{0.00085 \cdot 5} = 59$. To make matters worse, this is the case when we use the aperture stop radius to calculate the FN. However, as we need to accommodate the pupil wander off-axis, we obtain a much larger FN, whose magnitude depends on the stop distance. It is clear, therefore, that we cannot hope to obtain diffraction limited performance using these lens parameters. However, we can hope to make a significant improvement in performance compared to a chromatic metalens.

To evaluate the improvement our achromatic metalens provides over a standard chromatic metalens, we designed a competing chromatic metalens using cylindrical pillars. For the design of the chromatic metalens we chose for each radial position the pillar radius that most accurately achieved the required phase at the nominal wavelength, with no regard for the phase at other wavelengths. For the design of the achromatic metalens we searched the extended library for the nanostructure that most accurately achieved the required phase at 5 wavelengths, evenly spaced in the range 807nm-893nm, with varying weights [1, 1.2, 1.5, 1.2, 1].

Figure 15 shows the actual phase profiles obtained for the achromatic and chromatic metalens, based on their respective nanostructure libraries, compared to the theoretically desired phase profiles, which are given by Eq. (3):

$$\phi(r,\lambda) = \frac{\lambda_0}{\lambda}\phi(r,\lambda_0) + C(\lambda) = \frac{\lambda_0}{\lambda}\left(A_1 r^2 + A_2 r^4 + ...\right) + C(\lambda) \qquad (3)$$

Where $\lambda_0$ is the design wavelength (in our case 850nm), $A_i$ are the phase coefficients from Zemax, and $C(\lambda)$ is the previously mentioned phase bias. The $\lambda_0/\lambda$ factor comes from dividing the phase at the design wavelength, $\phi(r,\lambda_0)$, by $2\pi/\lambda_0$ to convert the phase to optical path difference (OPD) which should be constant for all wavelengths in an achromatic lens, and then multiplying by $2\pi/\lambda$ to obtain the corresponding phase at each wavelength.

The horizontal axis range shown in Figure 15 is only the central 0.25mm of the aperture radius, since after that the zones are too dense to visualize the phase. Near the center of the aperture the expanded nanostructure library of the achromatic metalens allows us to follow the ideal phase much more accurately for all the wavelengths. This improves not only the resolution of the metalens by correcting chromatic aberration, but also the efficiency, by better matching the zone profile to the ideal profile.

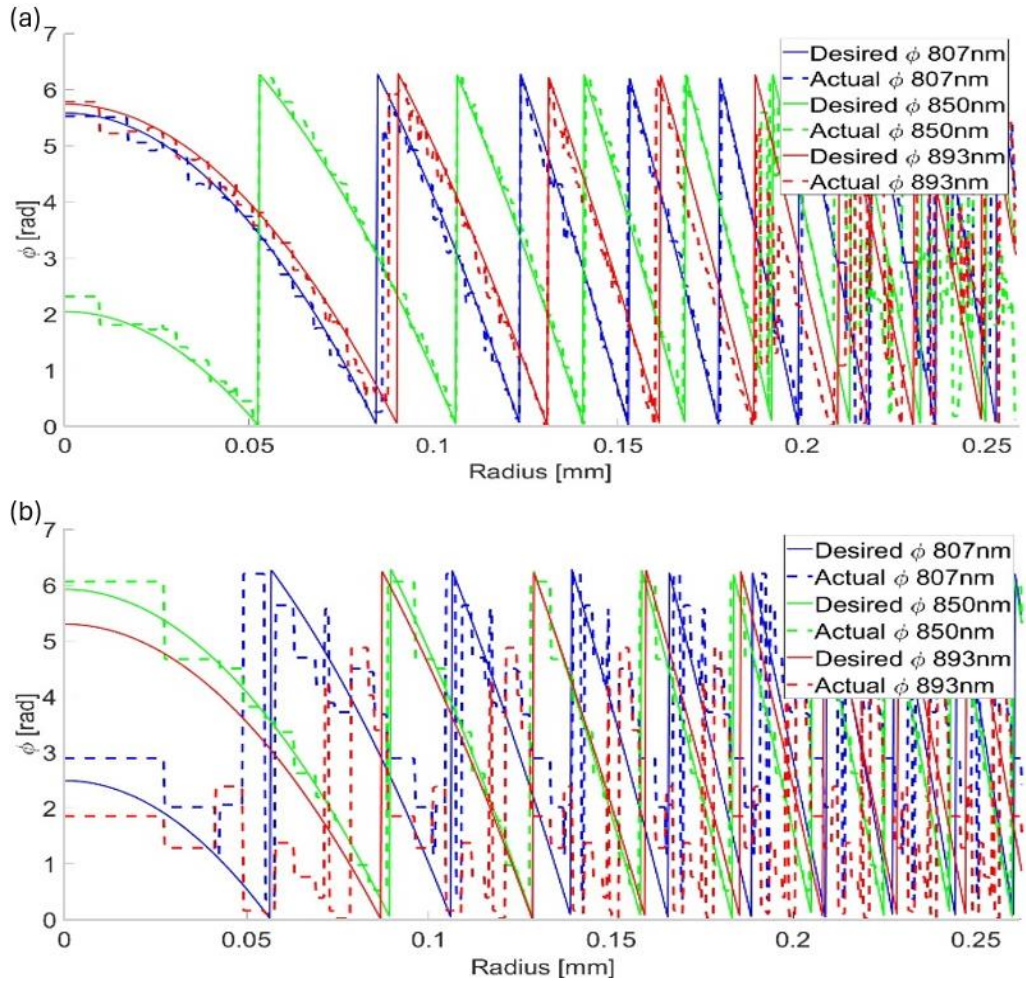

*Figure 15: Phase fitting at the center of the metalens for (a) achromatic metalens, and (b) chromatic metalens. Solid lines represent the ideal desired phase for each wavelength and dashed lines represent the phase that the chosen nanostructure achieves for each wavelength at each point along the aperture radius.*

# 7  Conclusion

In this paper we presented a nature inspired design method and rigorous performance analysis of a dispersion engineered achromatic wide-FOV metalens for the NIR spectral range. As such, the paper provides a recipe for the design of an achromatic metalens, and for properly evaluating its performance. Based on this approach we have designed a specific wide-FOV achromatic metalens with a fractional bandwidth of 10% operating in the NIR spectral range. Like the human visual system, our design provides high resolution on axis, at the expense of lower resolution in the peripheral FOV. This allows a significant improvement in performance compared to both conventional chromatic metalenses and previous achromatic designs. In addition to the improved resolution, we also achieved high efficiency of about 60% over the entire wavelength range. We hope that the insights from this paper, including the factors limiting the performance, and the use of rigorous methods of evaluation, will facilitate future advances in the field.

# 8 Acknowledgment

The research was partially funded by the Israeli Innovation authority, within the framework of the Meta-Materials and Meta-Surfaces consortium. R.M. Acknowledges a fellowship from the Center for Nanoscience and Nanotechnology, the Hebrew University of Jerusalem.

# 9 References

1. M. Khorasaninejad, W. T. Chen, R. C. Devlin, J. Oh, A. Y. Zhu, and F. Capasso, "Metalenses at visible wavelengths: Diffraction-limited focusing and subwavelength resolution imaging," Science (1979) **352**, 1190–1194 (2016).

2. P. Lalanne and P. Chavel, "Metalenses at visible wavelengths : past , present , perspectives," Laser Photonics Reviews **11**, 1600295 (2017).

3. J. Engelberg and U. Levy, "Optimizing the spectral range of diffractive metalenses for polychromatic imaging applications," Opt Express **25**, 21637–21651 (2017).

4. E. Arbabi, A. Arbabi, S. M. Kamali, Y. Horie, and A. Faraon, "Multiwavelength polarization-insensitive lenses based on dielectric metasurfaces with meta-molecules," Optica **3**, 628–633 (2016).

5. E. Arbabi, A. Arbabi, S. M. Kamali, Y. Horie, and A. Faraon, "Multiwavelength metasurfaces through spatial multiplexing," Sci Rep **6**, 1–8 (2016).

6. O. Avayu, E. Almeida, Y. Prior, and T. Ellenbogen, "Composite Functional Metasurfaces for Multispectral Achromatic Optics," Nat Commun **8**, 14992 (2017).

7. M. Khorasaninejad, Z. Shi, A. Y. Zhu, W. T. Chen, V. Sanjeev, A. Zaidi, and F. Capasso, "Achromatic Metalens over 60 nm Bandwidth in the Visible and Metalens with Reverse Chromatic Dispersion," Nano Lett **17**, 1819–1824 (2017).

8. E. Arbabi, A. Arbabi, S. M. Kamali, Y. Horie, and A. Faraon, "Controlling the sign of chromatic dispersion in diffractive optics," Optica **4**, 625–632 (2017).

9. W. T. Chen, A. Y. Zhu, V. Sanjeev, M. Khorasaninejad, Z. Shi, E. Lee, and F. Capasso, "A broadband achromatic metalens for focusing and imaging in the visible," Nat Nanotechnol **13**, 220–226 (2018).

10. S. Wang, P. C. Wu, V. C. Su, Y. C. Lai, M. K. Chen, H. Y. Kuo, B. H. Chen, Y. H. Chen, T. T. Huang, J. H. Wang, R. M. Lin, C. H. Kuan, T. Li, Z. Wang, S. Zhu, and D. P. Tsai, "A broadband achromatic metalens in the visible," Nat Nanotechnol **13**, 227–232 (2018).

11. S. Shrestha, A. C. Overvig, M. Lu, A. Stein, and N. Yu, "Broadband achromatic dielectric metalenses," Light Sci Appl **7**, 85 (2018).

12. F. Presutti and F. Monticone, "Focusing on Bandwidth: Achromatic Metalens Limits," Optica **7**, (2020).

13. J. Engelberg and U. Levy, "Achromatic flat lens performance limits," Optica **8**, 834–845 (2021).

14. A. Arbabi, E. Arbabi, S. M. Kamali, Y. Horie, S. Han, and A. Faraon, "Miniature optical planar camera based on a wide-angle metasurface doublet corrected for monochromatic aberrations," Nat Commun **7**, 13682 (2016).

15. J. Engelberg, C. Zhou, N. Mazurski, J. Bar-David, A. Kristensen, and U. Levy, "Near-IR wide field-of-view huygens metalens for outdoor imaging applications," Nanophotonics **9**, 361–370 (2020).

16. D. A. Buralli and G. M. Morris, "Design of a wide field diffractive landscape lens," Appl Opt **28**, 3950–3959 (1989).

17. J. Engelberg, C. Zhou, N. Mazurski, J. Bar-David, A. Kristensen, and U. Levy, "Near-IR wide-field-of-view Huygens metalens for outdoor imaging applications," Nanophotonics **9**, 361–370 (2020).

18. A. Martins, K. Li, J. Li, H. Liang, D. Conteduca, B. H. V. Borges, T. F. Krauss, and E. R. Martins, "On Metalenses with Arbitrarily Wide Field of View," ACS Photonics **7**, 2073–2079 (2020).

19. T. Xie, F. Zhang, M. Pu, H. Bao, J. Jin, J. Cai, L. Chen, I. Guo, X. Feng, Q. He, X. Ma, X. Li, B. Jiang, and X. Luo, "Ultrathin Wide-Angle and High-Resolution Meta-Imaging System via Rear-Position Wavevector Filter.pdf," Laser Photonics Rev. 2300119 (2023).

20. J. Engelberg and U. Levy, "Achromatic flat lens performance limits," Optica **8**, 834 (2021).

21. F. Yang, S. An, M. Y. Shalaginov, H. Zhang, C. Rivero-Baleine, J. Hu, and T. Gu, "Design of broadband and wide-field-of-view metalenses," Opt Lett **46**, 5735 (2021).

22. S. Luo, F. Zhang, X. Lu, T. Xie, M. Pu, Y. Guo, Y. Wang, and X. Luo, "Single-layer metalens for achromatic focusing with wide field of view in the visible range," J Phys D Appl Phys **55**, (2022).

23. J. Jang, G. Y. Lee, Y. Kim, C. Kim, Y. Jeong, and B. Lee, "Dispersion-Engineered Metasurface Doublet Design for Broadband and Wide-Angle Operation in the Visible Range," IEEE Photonics J **15**, 1–10 (2023).

24. Y. Hongli, C. Zhaofeng, and L. Xiaotong, "Broadband achromatic and wide field of view metalens-doublet by inverse design," Opt Express **32**, 15315 (2024).

25. J. Engelberg and U. Levy, "Generalized metric for broadband flat lens performance comparison," Nanophotonics **11**, 3559–3574 (2022).

26. J. W. Goodman, *Introduction to Fourier Optics*, 2nd ed. (McGraw-Hill, 1996).

27. J. Engelberg and U. Levy, "Standardizing flat lens characterization," Nat Photonics **16**, 171–173 (2022).

28. Z. Li, P. Lin, Y. W. Huang, J. S. Park, W. T. Chen, Z. Shi, C. W. Qiu, J. X. Cheng, and F. Capasso, "Meta-optics achieves RGB-achromatic focusing for virtual reality," Sci Adv **7**, 1–9 (2021).

29. M. Decker, W. T. Chen, T. Nobis, A. Y. Zhu, M. Khorasaninejad, Z. Bharwani, F. Capasso, and J. Petschulat, "Imaging Performance of Polarization-Insensitive Metalenses," ACS Photonics **6**, 1493–1499 (2019).

# Supporting information

# Nature Inspired Design Methodology for a Wide Field of View Achromatic Metalens

J. Engelberg,[1,2] R. Mazurski,[1] and U. Levy[1,*]

[1]Department of Applied Physics, The Faculty of Science, The Hebrew University of Jerusalem, Jerusalem, Israel, 9190401
[2]Department of Electro-optics and Applied Physics, Jerusalem College of Technology, Jerusalem, Israel, 9116001
*ulevy@mail.huji.ac.il

## 1 Zemax metalens prescriptions

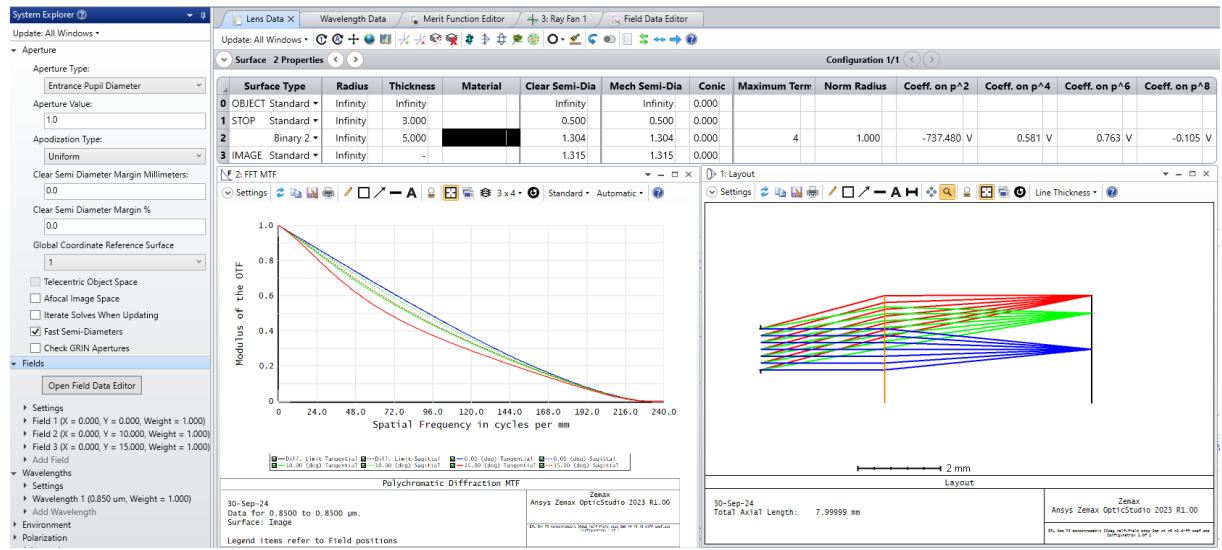

Figure S1: Metalens without substrate. Prescription and MTF performance, at 850nm single wavelength, 3mm stop distance, with high-order phase coefficients.

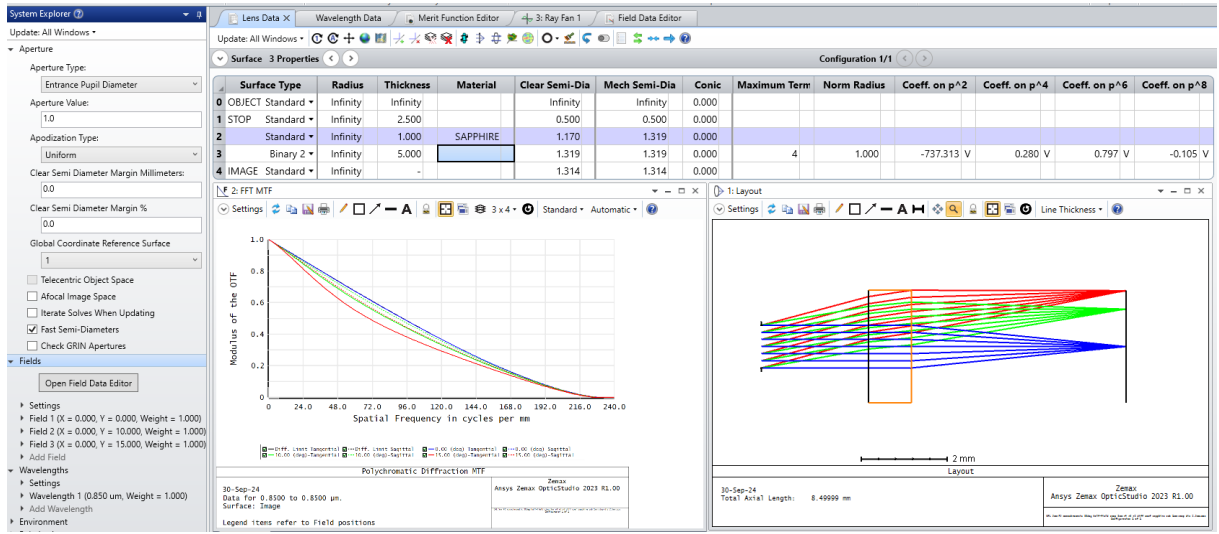

*Figure S2: Metalens with 1mm thick sapphire substrate. Prescription and MTF performance, at 850nm single wavelength, 2.5mm stop distance from substrate front surface (metalens is on rear surface), with high-order phase coefficients.*

The optical thickness of the 1mm substrate is given by:

$$t_{opt} = \frac{n-1}{n} t_{phys} = \frac{1.76 - 1}{1.76} \cdot 1 = 0.43 mm$$

Therefore, the overall equivalent stop distance for the design is 2.5+0.43=2.93mm. This is close to the 3mm design without substrate, which is why the two designs of Figure S1 and S2 give similar performance.

# 2  Achromatic metalens polychromatic performance at additional off-axis points

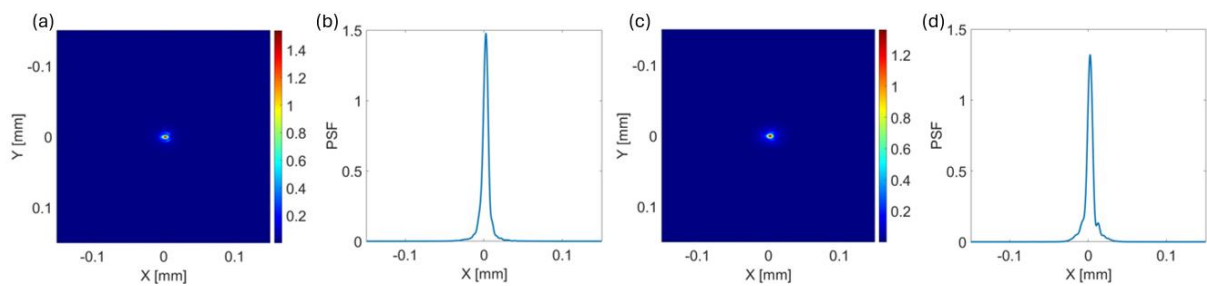

*Figure S3: 5° off-axis PSF comparison at stop distance of 2.5mm. (a-b) Achromatic metalens 2D PSF and cross-section in the radial (horizontal) direction respectively. (c-d) Chromatic metalens 2D PSF and cross-section in the radial (horizontal) direction respectively.*

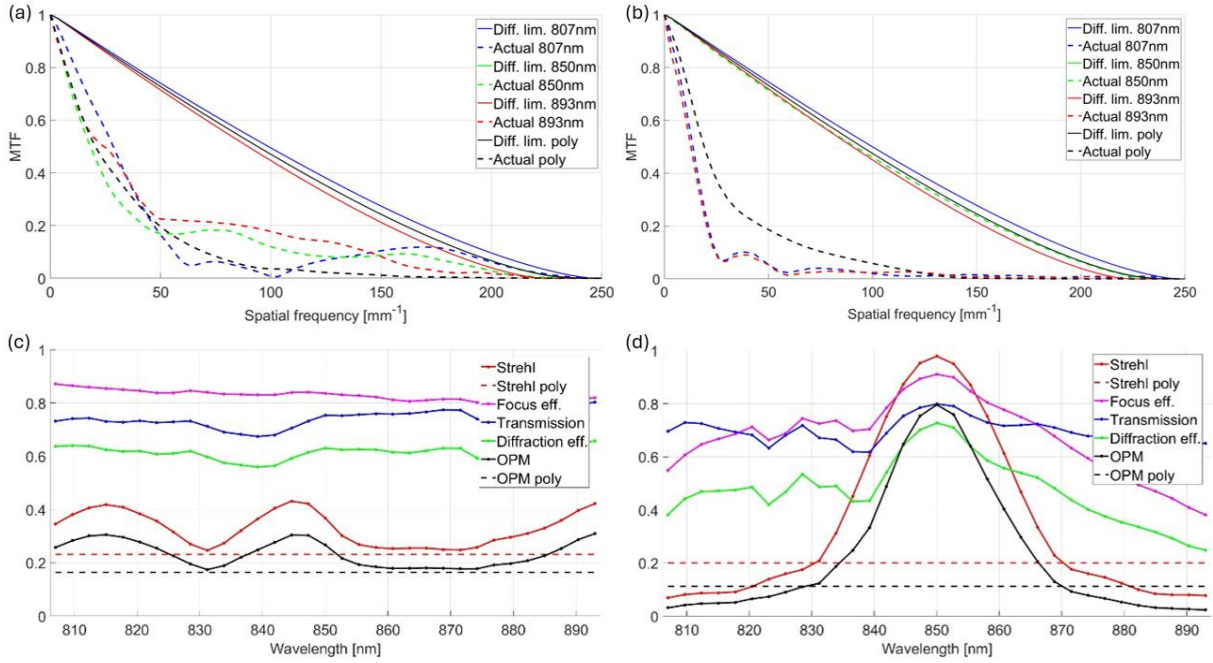

Figure S4: 5° off-axis performance comparison at stop distance of 2.5mm. (a) Achromatic metalens MTF. (b) Chromatic metalens MTF. (c) Achromatic metalens spectral and polychromatic Strehl, efficiency and overall performance (OPM) results as a function of wavelength. (d) The same for chromatic metalens.

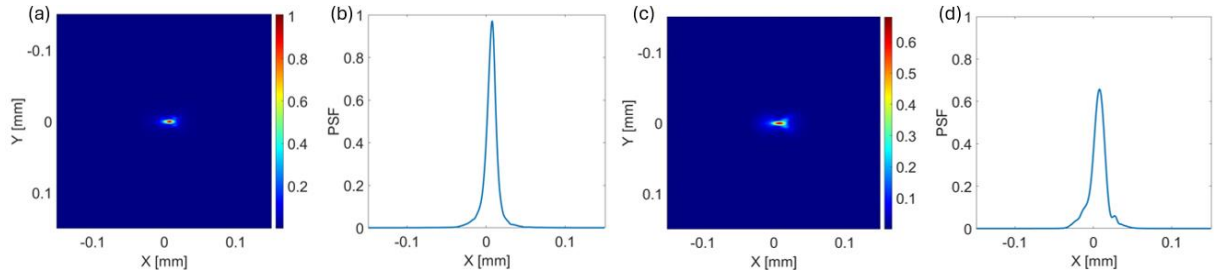

Figure S5: 10° off-axis PSF comparison at stop distance of 2.5mm. (a-b) Achromatic metalens 2D PSF and cross-section in the radial (horizontal) direction respectively. (c-d) Chromatic metalens 2D PSF and cross-section in the radial (horizontal) direction respectively.

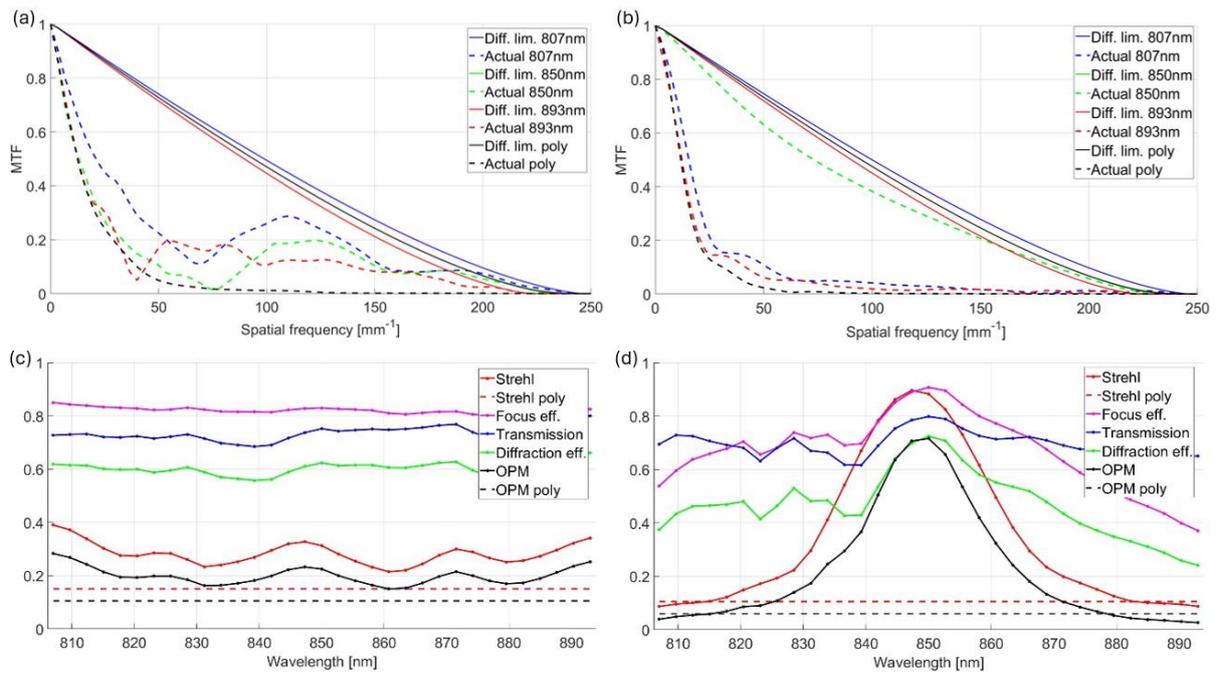

*Figure S6: 10° off-axis performance comparison at stop distance of 2.5mm. (a) Achromatic metalens MTF. (b) Chromatic metalens MTF. (c) Achromatic metalens spectral and polychromatic Strehl, efficiency and overall performance (OPM) results as a function of wavelength. (d) The same for chromatic metalens.*

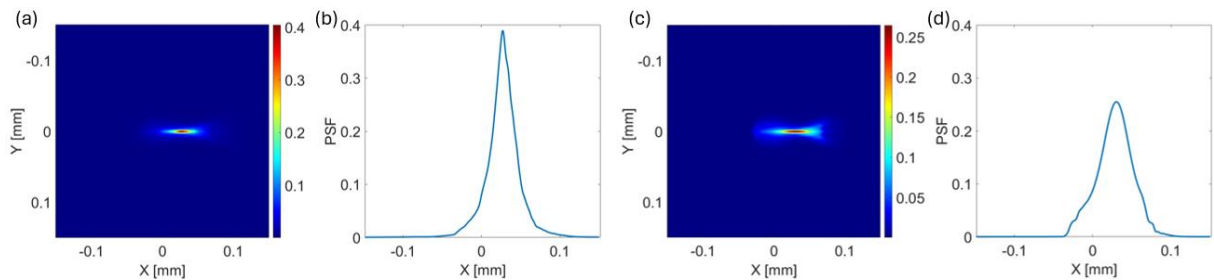

*Figure S7: 20° off-axis PSF comparison at stop distance of 2.5mm. (a-b) Achromatic metalens 2D PSF and cross-section in the radial (horizontal) direction respectively. (c-d) Chromatic metalens 2D PSF and cross-section in the radial (horizontal) direction respectively.*

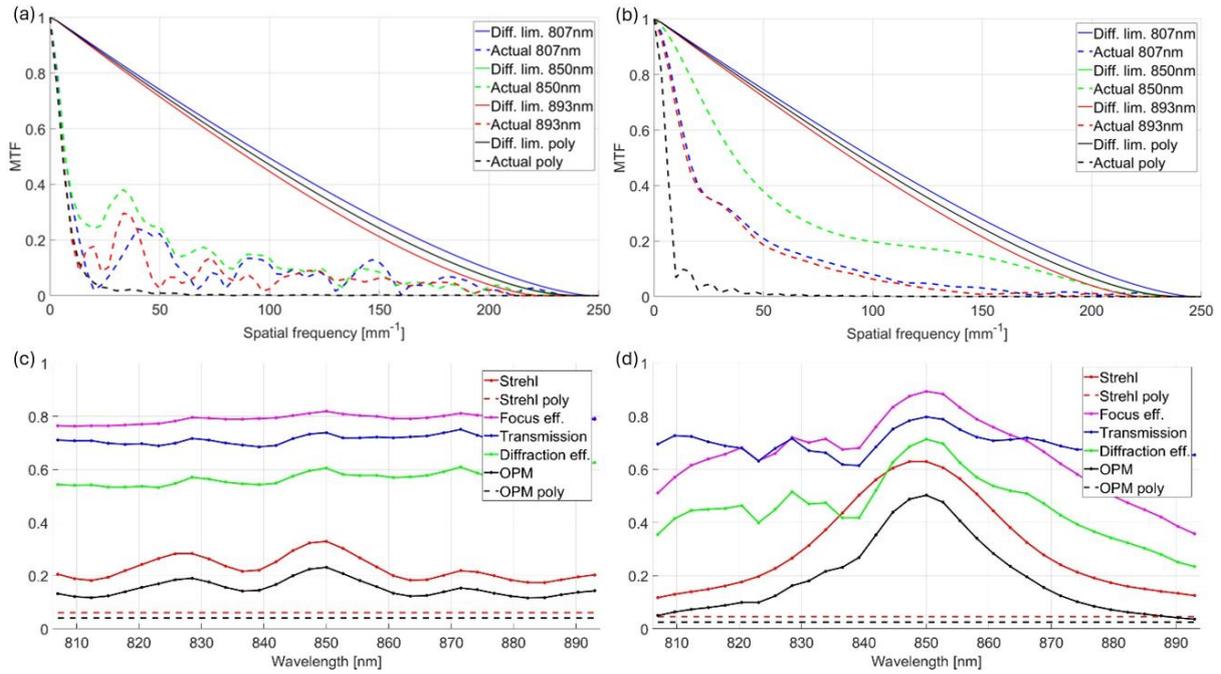

*Figure S8: 20° off-axis performance comparison at stop distance of 2.5mm. (a) Achromatic metalens MTF. (b) Chromatic metalens MTF. (c) Achromatic metalens spectral and polychromatic Strehl, efficiency and overall performance (OPM) results as a function of wavelength. (d) The same for chromatic metalens.*

The center of each PSF image, which is marked as $x = y = 0$, is located at a radial (horizontal in our images) distance of $f sin(\theta)$ relative to the optical axis. This is the chief ray height for the case of a quadratic phase metalens with stop at the front focal plane [1]. For this reason, when the aperture is not at the front focal plane and/or the phase is not quadratic, one can see that the center of the off-axis spot is not exactly on the center of the image. This does not affect the MTF results.

## 3   Achromatic metalens single wavelength performance

Following are images of PSFs at 3 single wavelengths (S9 – S13). It can be seen that the lateral chromatic aberration is partially corrected by the achromatic metalens by looking at off-axis PSFs (S10-S13). For the achromatic metalens the PSFs are approximately centered on the same *x* position (for each field angle) although they may seem distorted. Conversely, the chromatic metalens PSFs are shifted to the right for the short wavelength (807nm) and to the left for the long wavelength (893nm).

Another insight from the individual wavelengths PSFs is the ability to correct the axial chromatic aberration, as can be seen by looking at the peaks of the PSFs that are all on the same order of magnitude for the achromatic metalens, meaning similar Strehl ratio for all wavelengths. Conversely, for the chromatic metalens it is obvious that the enormous difference between the peaks of the design wavelength and the edge wavelengths is caused by defocusing of the latter.

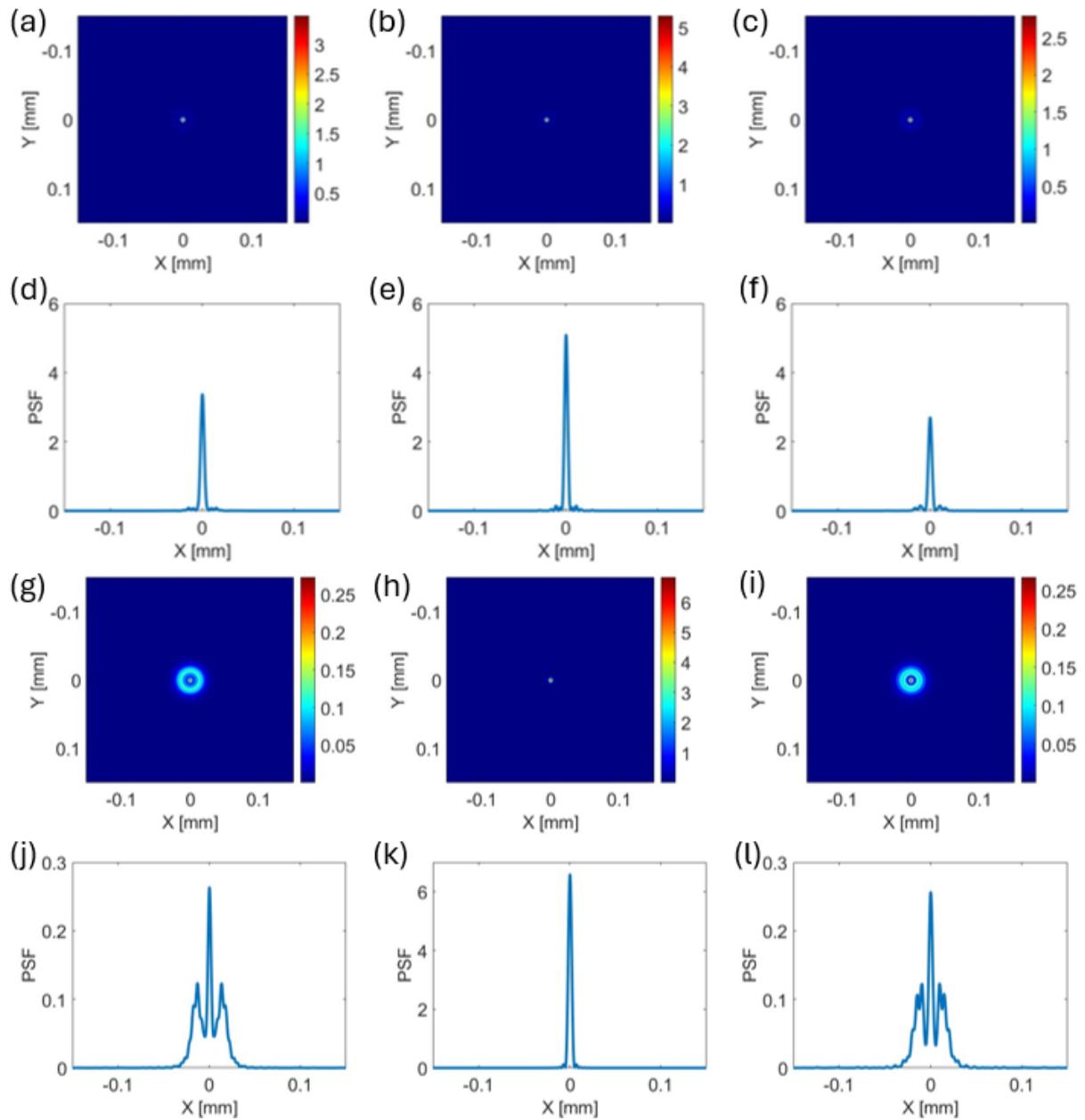

*Figure S9: On-axis PSF comparison for individual wavelengths. (a-f) Achromatic metalens. (g-l) Chromatic metalens. (a) and (g) are the PSFs at 807nm for achromatic and chromatic metalens, respectively. (d) and (j) are their cross sections, respectively. (b) and (h) are the PSFs at 850nm for achromatic and chromatic metalens, respectively. (e) and (k) are their cross sections, respectively. (c) and (i) are the PSFs at 893nm for achromatic and chromatic metalens, respectively. (f) and (l) are their cross sections, respectively.*

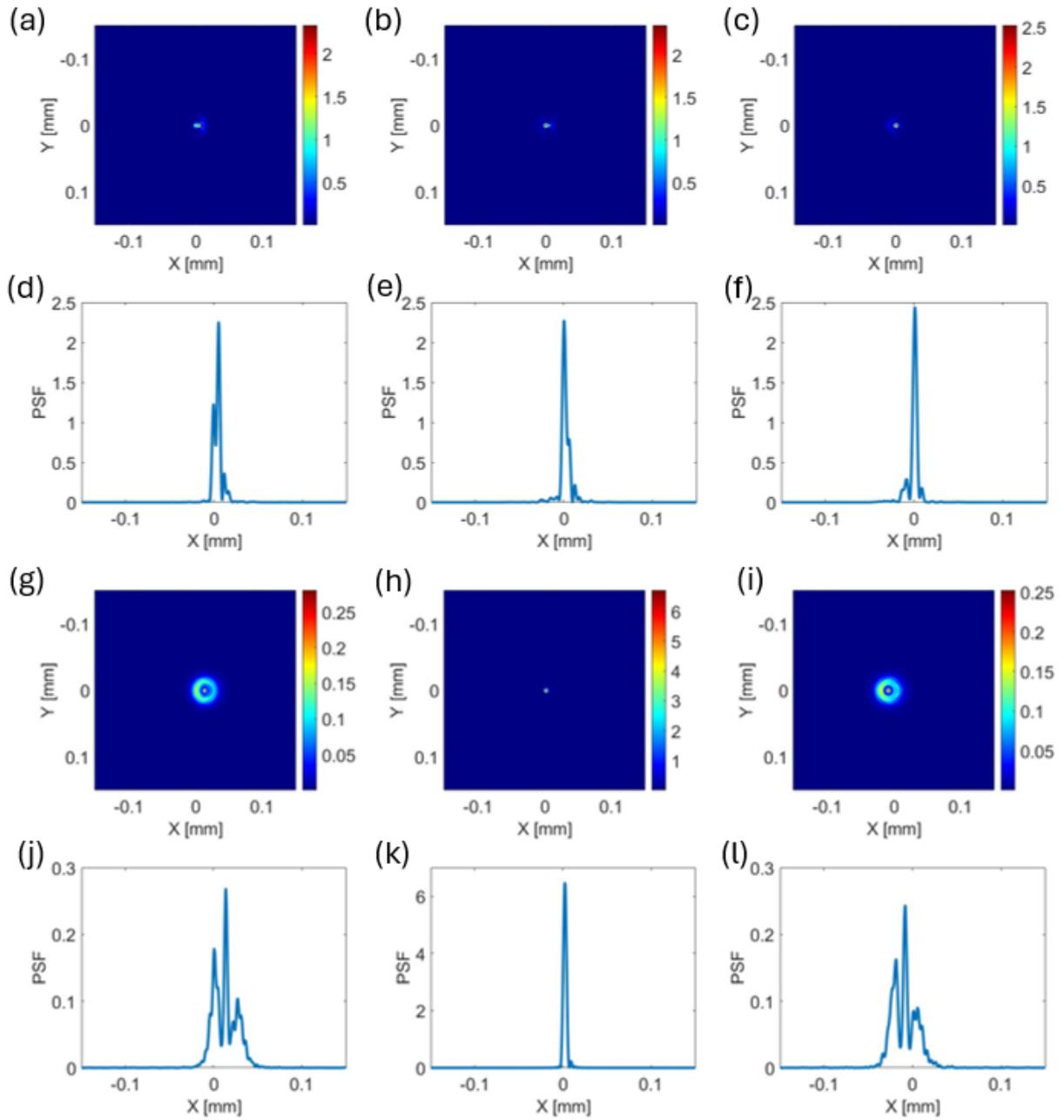

*Figure S10: 5˚ off-axis PSF comparison for individual wavelengths. (a-f) Achromatic metalens. (g-l) Chromatic metalens. (a) and (g) are the PSFs at 807nm for achromatic and chromatic metalens, respectively. (d) and (j) are their cross sections, respectively. (b) and (h) are the PSFs at 850nm for achromatic and chromatic metalens, respectively. (e) and (k) are their cross sections, respectively. (c) and (i) are the PSFs at 893nm for achromatic and chromatic metalens, respectively. (f) and (l) are their cross sections, respectively.*

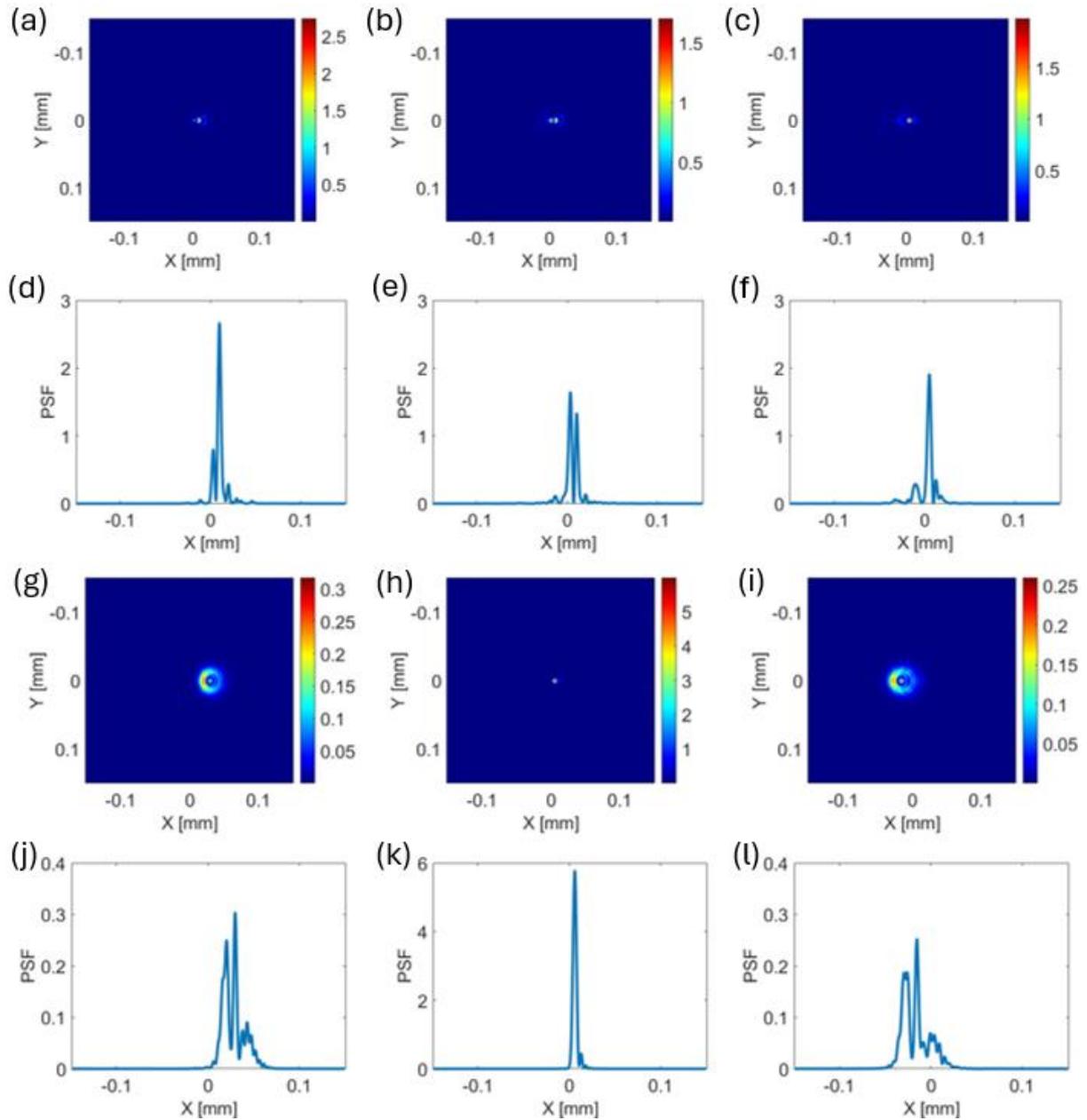

*Figure S11: 10° off-axis PSF comparison for individual wavelengths. (a-f) Achromatic metalens. (g-l) Chromatic metalens. (a) and (g) are the PSFs at 807nm for achromatic and chromatic metalens, respectively. (d) and (j) are their cross sections, respectively. (b) and (h) are the PSFs at 850nm for achromatic and chromatic metalens, respectively. (e) and (k) are their cross sections, respectively. (c) and (i) are the PSFs at 893nm for achromatic and chromatic metalens, respectively. (f) and (l) are their cross sections, respectively.*

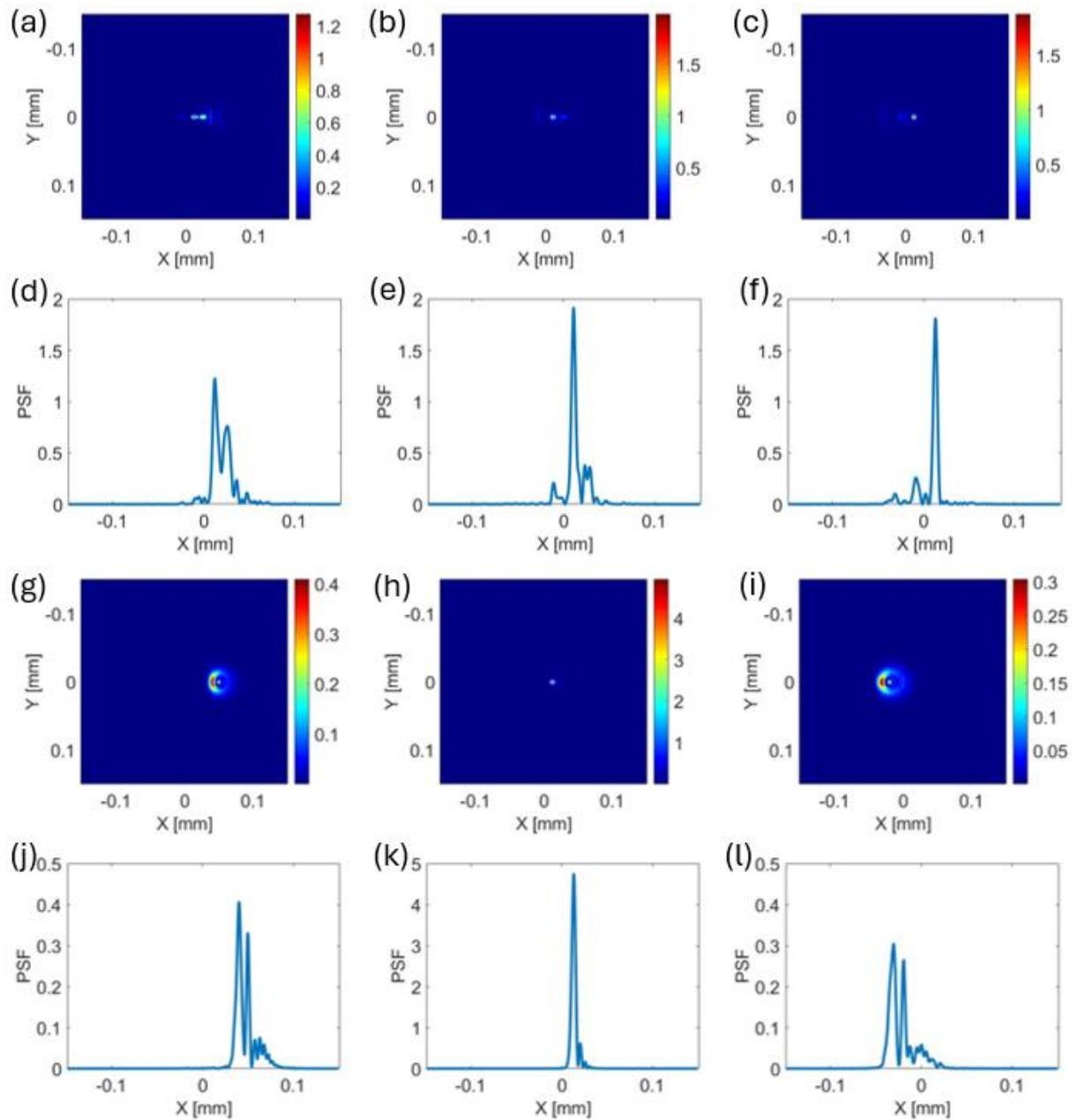

Figure S12: 15° off-axis PSF comparison for individual wavelengths. (a-f) Achromatic metalens. (g-l) Chromatic metalens. (a) and (g) are the PSFs at 807nm for achromatic and chromatic metalens, respectively. (d) and (j) are their cross sections, respectively. (b) and (h) are the PSFs at 850nm for achromatic and chromatic metalens, respectively. (e) and (k) are their cross sections, respectively. (c) and (i) are the PSFs at 893nm for achromatic and chromatic metalens, respectively. (f) and (l) are their cross sections, respectively.

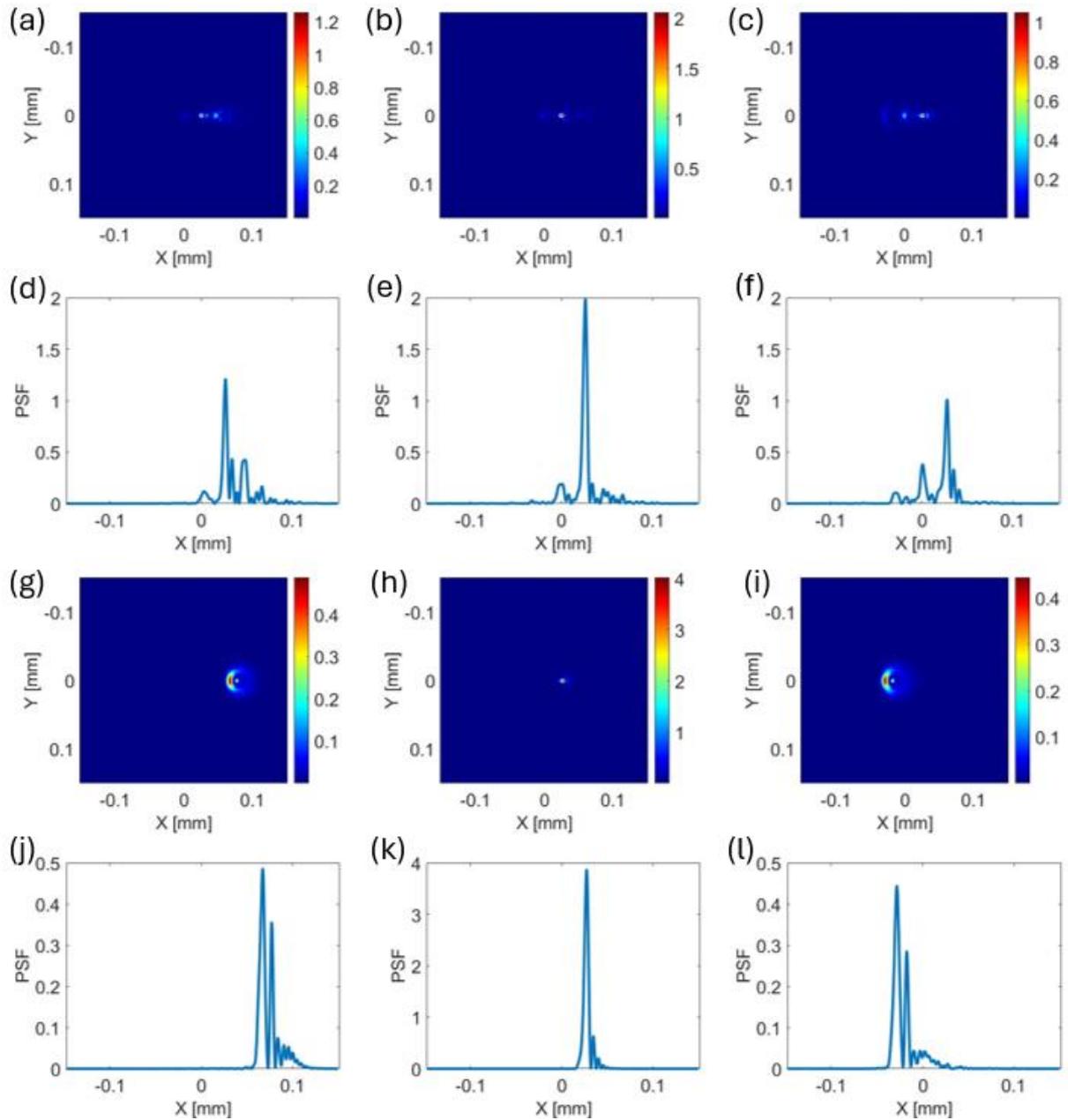

*Figure S13: 20° off-axis PSF comparison for individual wavelengths. (a-f) Achromatic metalens. (g-l) Chromatic metalens. (a) and (g) are the PSFs at 807nm for achromatic and chromatic metalens, respectively. (d) and (j) are their cross sections, respectively. (b) and (h) are the PSFs at 850nm for achromatic and chromatic metalens, respectively. (e) and (k) are their cross sections, respectively. (c) and (i) are the PSFs at 893nm for achromatic and chromatic metalens, respectively. (f) and (l) are their cross sections, respectively.*